\documentclass[12pt]{article}
\usepackage{amsmath}
\usepackage{graphicx,psfrag,epsf,subcaption}
\usepackage{enumerate}
\usepackage{natbib}
\usepackage{url} 
\usepackage{amssymb}
\usepackage{bm}
\usepackage{tikz}
\usepackage{multirow}
\usepackage{float}
\usepackage{rotating}
\usepackage{hyperref}
\usepackage{geometry}
\usepackage{pdflscape}
\usepackage{array}
\usepackage{booktabs}

\newcommand{\blind}{0}

\newcommand{\numbereqn}{\addtocounter{equation}{1}\tag{\theequation}} 

\newcolumntype{L}{>{\centering\arraybackslash}m{3.6cm}}

\makeatletter
\newcommand{\thickhline}{%
    \noalign {\ifnum 0=`}\fi \hrule height 1.8pt
    \futurelet \reserved@a \@xhline
}
\newcolumntype{"}{@{\hskip\tabcolsep\vrule width 1.5pt\hskip\tabcolsep}}
\makeatother

\def\revise{\color{black}}

\begin{document}

\def\spacingset#1{\renewcommand{\baselinestretch}%
{#1}\small\normalsize} \spacingset{1}


\if0\blind
{
  \title{\bf {A Unified Three-State Model Framework for Analysis of Treatment Crossover in Survival Trials}}
  \author{Zile Zhao$^a$, Ye Li$^b$, Xiaodong Luo$^b$, Ray Bai$^a$\thanks{Corresponding author: rbai@mailbox.sc.edu. The authors gratefully acknowledge financial support from the McCausland Innovation Fund and the University of South Carolina Office of the Vice President for Research ASPIRE program.}  \vspace{0.2cm}\\ 
    $^a$Department of Statistics, University of South Carolina, Columbia, SC 29208 \\ 
    $^b$Evidence Generation \& Decision Science, R\&D, Sanofi US, Bridgewater, NJ 08807 \\ }
  \maketitle
} \fi

\if1\blind
{
  \bigskip
  \bigskip
  \bigskip
  \begin{center}
    {\LARGE\bf Title}
\end{center}
  \medskip
} \fi

\bigskip
\begin{abstract}
We present a unified three-state model (TSM) framework for evaluating treatment effects in clinical trials in the presence of treatment crossover. Researchers have proposed diverse methodologies to estimate the treatment effect that would have hypothetically been observed if treatment crossover had not occurred. However, there is little work on understanding the connections between these different approaches from a statistical point of view. {\revise The} proposed TSM framework unifies existing methods, effectively identifying potential biases, model assumptions, and inherent limitations for each method. This can guide researchers in understanding when these methods are appropriate and choosing a suitable approach for their data. The TSM framework also facilitates the creation of new methods to adjust for confounding effects from treatment crossover. To illustrate this capability, we introduce a new imputation method that falls under its scope. Through simulation experiments, we demonstrate the performance of different approaches for estimating the treatment effects. {\revise Codes for implementing the methods within the TSM framework are available at \url{https://github.com/JasonZhao111/TSM}.}
\end{abstract}

\noindent%
\vfill

\spacingset{1.45} 
\section{Introduction}
\label{sec:intro}
Randomized controlled trials (RCTs) have been commonly used to compare the survival outcomes between a treatment group and a control group. Treatment crossover within an RCT takes place when patients randomized to a treatment or control group switch to a different group from the one to which they were initially assigned \citep{Ishak2014, Linus2014, Watkins2013}. {\revise We refer to these individuals as ``crossed-over patients".}

Treatment crossover can occur for a variety of reasons and can be either noninformative or informative. We define noninformative crossover as any scenario where patients  switch treatments for reasons unrelated to treatment efficacy, disease progression, or any meaningful clinical factors. An example of noninformative crossover is a study protocol which allows patients in the control group to switch to the experimental treatment once the main study endpoint (e.g. the blinded phase of the trial) has been reached \citep{Daugherty2008}. 
For example, in the CENTAUR trial of sodium phenylbutyrate and taurursodio (PB and TURSO) in amyotrophic lateral sclerosis (ALS), patients in the control group all switched to the experimental treatment after they completed the 24-week double-blind study phase \citep{Paganoni2022}. 

However, treatment crossover can also be informative. Informative crossover occurs when patients switch treatments as a result of disease progression \citep{Ishak2014, Latimer2014}.  Informative treatment crossover is often a {\revise selection} process. For example, only patients who are deemed likely to benefit from the experimental treatment may be crossed over, while those who have reached the terminal stage of the disease typically are not \citep{Ishak2014}. In many real-life scenarios, crossover can also result from a \emph{mixture} of both noninformative and informative events, such as the other treatment arm being perceived to be better \emph{and} a sudden change in the patient's condition \citep{Ishak2011}. 

\begin{figure}[t!]
	\centering
\includegraphics[scale=0.66]{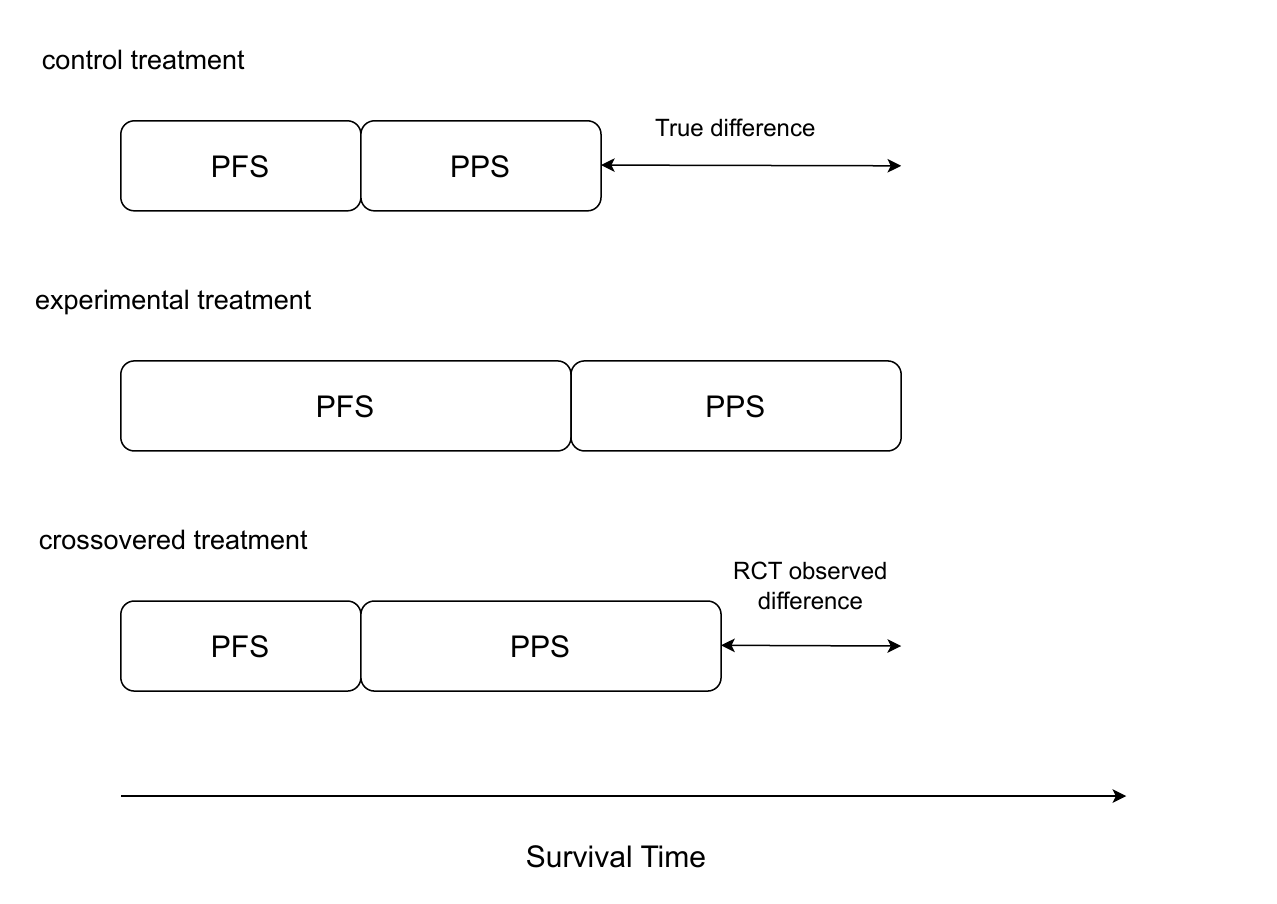}
	\caption{Diagram of PFS and PPS for patients assigned to the control, experimental, and {\revise crossed-over} treatments}
	\label{fig:PFS}
\end{figure}

These diverse scenarios invariably introduce complications in data analysis to estimate the treatment efficacy. Under treatment crossover, conventional statistical techniques such as intention-to-treat analysis \citep{Brody2016} (described in Section \ref{sec:TSM_methods}) may yield biased estimations of the true treatment effect. 
To illustrate this, we consider the following example. Progression-free survival (PFS) is a commonly used clinical endpoint, particularly in Oncology \citep{Dancey2009}. PFS measures the duration between randomization and disease progression or death. Post-progression survival (PPS) refers to the time \emph{after} disease progression to death. Figure \ref{fig:PFS} shows that the survival time for a {\revise crossed-over} patient represents a mixed effect of having received both the control and experimental treatments. The observed difference in survival time is biased as a result of this {\revise mixture}, and the true treatment effect is underestimated \citep{Ishak2014, Latimer2014}.

An intuitive approach to account for treatment crossover is to exclude or censor data at the point of crossover. This is also known as the per-protocol method \citep{Brody2016}, where researchers only focus on PFS when crossover is linked with the disease progression. However, the per-protocol method can {\revise also} be vulnerable to selection bias, as individuals who switched from the control treatment to the experimental treatment might have different prognoses and therefore are not exchangeable with patients who stayed in the control group \citep{ArnoldErcumen2016}.  Potential solutions such as randomized crossover designs can be applied to mitigate the effects of selection bias \citep{LiamMcKeever2021, SimonChinchilli2007}. However, randomization -- while effective in theory -- is not always practical or ethical \citep{SimonChinchilli2007}. A more pragmatic approach may be to \emph{adjust} for crossover effects \emph{within} the statistical analysis.

Numerous approaches for estimating the true treatment effect (or hazard ratio) in the presence of treatment crossover have been proposed and examined. In Section \ref{sec:TSM_methods}, we describe several of the most popular methods. \cite{Morden2011} and \cite{Latimer2018} conducted simulations in a variety of plausible crossover scenario settings and compared the results from these different methods. Other recent reviews covering a range of therapeutic areas can be found in \citet{Watkins2013}, \citet{Ishak2014}, and \citet{Linus2014}. 

Despite these reviews and comparative studies, there is little work on understanding \emph{how} these different methods relate to each other in terms of their statistical properties. While assessing these methods' performance under various simulation settings \citep{Morden2011, Latimer2018} is certainly illuminating, it is not always straightforward for practitioners to know which statistical method they should use to adjust for confounding from treatment crossover. Moreover, based on domain knowledge or evidence from past trials, a researcher may also want to incorporate their own assumptions (e.g. an assumption that the treatment effect is constant over time) into their statistical analyses \citep{KahanMorris2013, Glidden2020}. 

In this paper, we propose to use a three-state model (TSM) framework to synthesize existing statistical methods in treatment crossover analysis. In addition to unifying existing methods under one statistical umbrella, {\revise the} TSM framework easily enables the creation of \emph{new} methods allowing the incorporation of diverse assumptions tailored to specific scenarios. We demonstrate the utility of the TSM framework by introducing a novel method for estimating the {\revise treatment effect}. We find that all methods rely on critical limiting assumptions, and the accuracy of estimating the treatment effect relies on the validity of these assumptions. Our framework can guide practitioners in determining (or inventing) the most appropriate method to use for analyzing their data from clinical trials with treatment crossover.


The rest of this paper is structured as follows. Section \ref{Sec:TSM} formally introduces the TSM framework. 
Section \ref{sec:TSM_methods} places existing methods in the context of this framework and introduces a new imputation method. Section \ref{Sec:sims} demonstrates the methodologies through simulation experiments. Section \ref{Sec:Discussion} concludes the paper with a brief discussion.

\section{Three-State Model Framework for Treatment Crossover} \label{Sec:TSM}

The main thrust of this article is to introduce a broad statistical framework for understanding the assumptions, limitations, and relationships of different methods for estimating treatment effects in RCTs with treatment crossover. We first introduce the TSM framework and discuss treatment effect estimation under this framework.

\subsection{Statistical Framework}

In an RCT, we define the \emph{entry} as the time that a patient enters the trial. Let $T$ be the time from entry to the event of interest (e.g. death), and let $U$ be the time from entry to the time of crossover.  Patients can arrive at the event of interest {\revise via two different paths as illustrated in Figure \ref{fig:TSM_diagram}}. The first {\revise path} is to go to the event directly from entry without crossover, characterized by a hazard function $\lambda_{1}(t)$. The second {\revise path} is where {\revise the crossover event} (e.g. re-randomization, progression, or end of the double-blind phase) occurs. {\revise For this path}, we let $\lambda_{2}^{\bf x}(t \mid u)$ denote the hazard function from crossover to the event. The crossover path is completed with {\revise a} path connecting the entry point and the crossover point, and we represent the hazard function for this path as $\lambda_{3}(u)$. Formally, the three hazard functions are defined as follows:

\begin{figure}[t!]
  \centering
  \includegraphics[width = .82\textwidth]{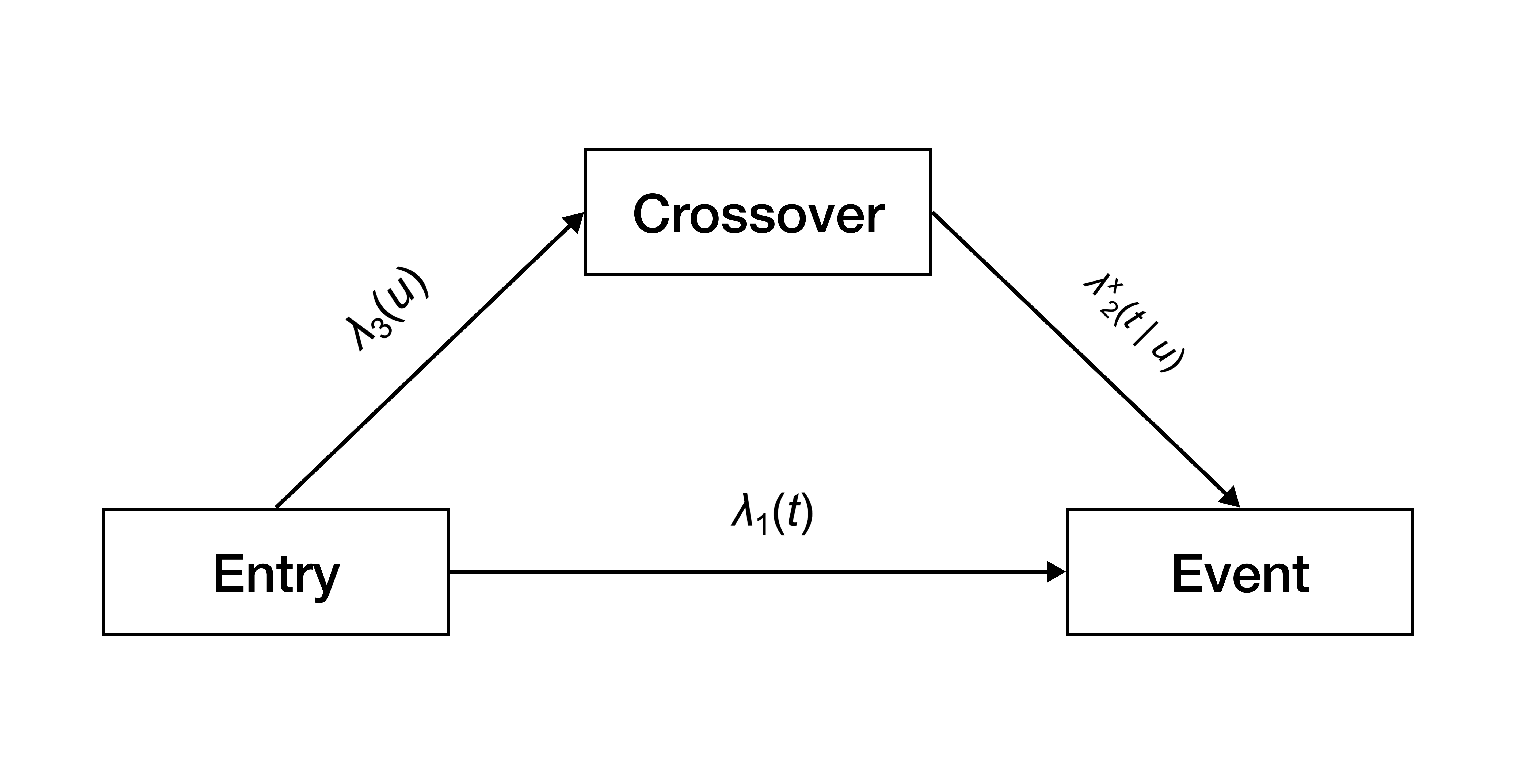}
  \caption{Diagram of the three-state model}
  \label{fig:TSM_diagram}
\end{figure}

\begin{description}
\item[(1) hazard from entry to event without crossover:]
\begin{equation} \label{lambda1}
    \lambda_1(t)= P(T=t\mid T\ge t,U\ge t), \hspace{0.5cm}t>0.
\end{equation}

\item[(2) hazard from crossover to event (if there is crossover):]
\begin{equation} \label{lambda2}
\lambda^{\bf x}_{2}(t\mid u)= P(T=t\mid T\ge t,U=u),\hspace{0.5cm}t>u.
\end{equation}

\item[(3) hazard from entry to crossover:]
\begin{equation} \label{lambda3}
    \lambda_3(u)= P(U
=u\mid T\ge u,U\ge u), \hspace{0.5cm}u>0.
\end{equation}
\end{description}
{\revise This multi-state model has been studied and used for survival analysis \citep{Meira-Machado2009, xu2010} and for designing clinical trials allowing for treatment crossover \citep{xia2016, xiaodong2019}. In this paper, we use this multi-state model to establish a framework to unify different methods for handling treatment crossover. First we assume} 
$U =\infty$ when $U >T$, {\revise i.e.,} crossover occurring after the event cannot be observed or is of no interest. {\revise With this assumption, the joint distribution of $U$ and $T$ is uniquely determined by the hazard functions $\lambda_1(t)$, $\lambda^{\bf x}_{2}(t\mid u)$, and $\lambda_3(u)$ \citep{xu2010}. Note that this model setting does not assume independence between $U$ and $T$.}  
{\revise The focus is to evaluate the impact of treatment crossover on} the survival function $\widetilde{S}(t)= P(T>t)$ {\revise when different methods are used}. Routine calculations {\revise \citep{Meira-Machado2009}} give
\begin{align*} \label{Stilde}
\widetilde{S}(t) &= P(T>t,U >t)+P(T>t\ge U)\\
     &= S_1(t)S_3(t)+\int_0^t P(T>t, T\ge u, U=u)du\nonumber\\
     &= S_1(t)S_3(t)+\int_0^t  \exp\Big\{-\int_u^t \lambda^{\bf x}_{2}(s\mid u)ds\Big\} \lambda_3(u)S_1(u)S_3(u)du, \numbereqn
\end{align*}
where $S_{j}(t)=\exp\{-\int_0^t \lambda_j(s)ds\}$, $j=1,3$.  We can correspondingly define the density function $\widetilde{f}(t)={\revise -d\widetilde{S}(t)/dt}$, the cumulative hazard function $\widetilde{\Lambda}(t)=-\log \{\widetilde{S}(t)\}$, and the hazard function $\widetilde{\lambda}(t)$,
\begin{equation} \label{lambdatilde}
\widetilde{\lambda}(t) = \widetilde{f}(t)/\widetilde{S}(t).
\end{equation}
{\revise In the special case when there is no crossover, i.e., $\lambda_3(u)=0$ for all $u>0$, $\widetilde{S}(t)=S_1(t)$. Otherwise, $\widetilde{S}(t)$,}
$\widetilde{f}(t)$, $\widetilde{\Lambda}(t)$, and $\widetilde{\lambda}(t)$ depend on the three hazard functions $\lambda_1$, $\lambda_2^{\bf x}$, and $\lambda_3$ defined in \eqref{lambda1}-\eqref{lambda3}. Various configurations of $\lambda_2^{\bf x}$ result in distinct forms of crossover. In what follows, {\revise $\lambda(\cdot)$} denotes a generic hazard function. We focus on the following two types of crossover which are frequently observed: 
\begin{enumerate}
\item \textbf{Markov crossover}. The hazard function after crossover only depends on the time from entry to event, {\revise i.e.,} 
\begin{align} \label{Markov_assumption}
    \lambda_2^{\bf x}(t \mid u) =\lambda(t). 
\end{align}
\citet{xiaodong2019} termed this as Markov crossover {\revise and provided some examples for which this type of crossover is considered reasonable.}
\item \textbf{Semi-Markov crossover}. The hazard function after crossover depends only on the duration of time between the crossover point and the event, {\revise i.e.,} 
\begin{align} \label{semi-Markov_assumption}
\lambda_2^{\bf x}(t \mid u)=\lambda(t-u).
\end{align}
In this case, patients will experience a new hazard function beginning from the crossover point. {\revise \cite{xiaodong2019}} noted that semi-Markov crossover is frequently observed as a result of disease progression and/or re-randomization.
\end{enumerate}
{\revise Note that the hazard function $\lambda(\cdot)$ in \eqref{Markov_assumption} and \eqref{semi-Markov_assumption} can take different forms. The difference between Markov and semi-Markov crossovers lies in whether this function applies to the time from study entry or the time from crossover. There are other types of crossover. For example, \citet{xiaodong2019} introduced a hybrid form of Markov and semi-Markov crossover. More generally, $\lambda_2^{\bf x}(t \mid u)$ can be any non-negative bivariate function of $t$ and $u$. However, Markov and semi-Markov crossovers are most commonly seen and are easier to handle. As elaborated in Section \ref{sec:TSM_methods}, existing methods for handling treatment crossover assume that the crossover is either Markov or semi-Markov. }

\subsection{Treatment Effect}

Let $A$ {\revise denote} the control group and $B$ {\revise denote} the treatment group. Without loss of generality, we will focus on the case where crossover can only occur in the control patients, and the {\revise crossed-over} patients begin their new treatment from the time of crossover. We begin by establishing the following notation, where the hazard function is defined as in \eqref{lambdatilde}:
\begin{itemize}
\item $\widetilde{\lambda}_A(t)$: the overall hazard function for the treatment group;
\item $\widetilde{\lambda}_B(t)$: the overall hazard function for the control group patients if crossover to treatment is \emph{not} allowed;
\item $\widetilde{\lambda}_{B*}(t)$: the  
hazard function for the control group patients if crossover to treatment is allowed.
\end{itemize}
The true treatment efficacy can be quantified by comparing $\widetilde{\lambda}_A(t)$ with $\widetilde{\lambda}_B(t)$, rather than with $\widetilde{\lambda}_{B*}(t)$. The treatment effect we seek to estimate is therefore the \emph{hazard ratio} (HR),
\begin{equation} \label{treatmenteffect}
\text{HR}(t) = \widetilde{\lambda}_A(t) / \widetilde{\lambda}_B(t).
\end{equation}
An $\mbox{HR}(t)$ exactly equal to one indicates equal efficacy of the experimental and control treatments. On the other hand, an $\mbox{HR}(t)$ less than one favors the experimental treatment, while an $\mbox{HR}(t)$ greater than one favors the control. 

In {\revise the} TSM framework, we have $\widetilde{\lambda}_{B*}(t) = H(t; \lambda_1,\lambda_{2*},\lambda_3)$ and $\widetilde{\lambda}_{B}(t) = H(t; \lambda_1,\lambda_{2},\lambda_3)$, where $H$ denotes a function of three hazards, $\lambda_{2*}$ denotes the hazard function if {\revise the} crossover {\revise event} occurs \emph{and} the control patient {\revise \emph{switches}} to the experimental treatment, and $\lambda_{2}$ is the hazard function if the crossover {\revise event} occurs but the control patient \emph{remains} in the control group.

Let $c$ be an indicator for whether the patient switches to the experimental treatment after {\revise the} crossover {\revise event} ($c=1$ if ``yes'', $c=0$ if ``no''). The hazard functions $\lambda_{2*}$ and $\lambda_{2}$ have the forms,
\begin{align} \label{lambda2star_and_lambda2}
    \begin{array}{rl}
        \lambda_{2*}(t \mid u) &= \lambda_{2}^{\bf x} (t \mid u ,c=1),   \\
        \lambda_{2}(t \mid u) &= \lambda_{2}^{\bf x} (t \mid u , c=0),
    \end{array}
\end{align}
where $\lambda_2^{\bf x}$ is defined as in \eqref{lambda2}.
In the case where a proportion of the patients remains in the control group after the crossover {\revise event}, we can estimate $\lambda_{2}$ in \eqref{lambda2star_and_lambda2} using various statistical methods. However, if all patients switch to the treatment after the crossover {\revise event},  $\lambda_{2}$ will become non-identifiable because there is no data on patients continuing the control treatment. In this case, additional assumptions are required to ensure identifiability.

\section{Methods Within the TSM Framework} \label{sec:TSM_methods}

{\revise The} TSM framework subsumes many existing methods for estimating treatment effects \eqref{treatmenteffect} in the presence of treatment crossover. Furthermore, new methods can also be introduced under this framework. In this section, we synthesize several existing methods that fall under the TSM umbrella and introduce a new imputation method as part of the TSM toolkit. Table \ref{tab:summary} provides a summary of all the methods we consider.

The best method(s) to use in practice depends on the validity or plausibility of the researchers' assumptions. If Markov crossover \eqref{Markov_assumption} can be reasonably assumed, then one of the Markov models in Table \ref{tab:summary} can be used.  On the other hand, if semi-Markov crossover \eqref{semi-Markov_assumption} is more reasonable, then one of the semi-Markov models in Table \ref{tab:summary} can be pursued. When crossover is linked specifically to re-randomization or disease progression, the semi-Markov assumption \eqref{semi-Markov_assumption} is especially plausible \citep{xiaodong2019}. Within each of these model classes, further assumptions can be imposed depending on the specific circumstances of the RCT.

\subsection{Intention-to-treat} \label{sec:ITT_method}
Numerous authors adopt a practical approach by applying an intention-to-treat (ITT) analysis. In the ITT method, analysis of patients is based solely on the treatment group to which they were originally randomized, regardless of their adherence with the entry criteria and regardless of protocol deviation or participant withdrawal \citep{Brody2016}. The fundamental crux of an ITT analysis is to use the complete dataset of patients who were subjected to randomization at the very beginning of the RCT \citep{Brody2016}.

ITT analysis results should be reported in all cases. This is essential because ITT upholds the integrity of randomized trials by analyzing participants according to their original assigned groups, regardless of their adherence or completion of the allocated treatment \citep{Brody2016}. Although the ITT analysis is generally acceptable, it potentially underestimates the true policy effectiveness of a treatment \citep{Ian2005}. For example, in fatal conditions like ALS, it is often recommended to allow patients to switch to experimental treatments at a certain time point \citep{Paganoni2022}. Therefore the estimated treatment effect from the ITT analysis may be diluted \citep{Ian2005, Paganoni2022}.

In the TSM framework, ITT actually compares $\widetilde{\lambda}_{A}(t)$ with $\widetilde{\lambda}_{B*}(t)$ instead of with $\widetilde{\lambda}_{B}(t)$ as in \eqref{treatmenteffect}. In other words, ITT does not make any adjustment for treatment crossover, and it maintains the most conservative approach in treatment crossover analysis.

\begin{table}[t!]
\centering
\caption{Table of different methods falling under the TSM framework} \label{tab:summary}
\resizebox{\textwidth}{!}{
\begin{tabular}{|l|c|L|L|}   
\hline
\textbf{Method} & \textbf{Abbreviation}   & \textbf{Assumed crossover type} & \textbf{Use data after crossover?} \\
\thickhline 
Intention-to-treat & ITT  & N/A            & Yes               \\
Censor-at-switching & CAS & Markov            & No\\
Exclude-at-switching & EAS & Markov            & No  \\
Treatment as time-dependent variable & TTDV & Markov            & Yes \\
Rank preserving structural failure time & RPSFT  & Semi-Markov & Yes \\
{\revise Two-stage accelerated failure time} & {\revise TSAFT}  & {\revise Semi-Markov} & {\revise Yes} \\
Inverse-probability-of-censoring weighting &  IPCW & N/A  & No \\
Bayesian imputed multiplicative method & BIMM  & Semi-Markov    & Yes\\
\hline
\end{tabular}}
\end{table}

\subsection{Per-Protocol} \label{sec:PP}
A per-protocol (PP) analysis entails evaluating patients based only on the treatment they were actually administered, rather than the one to which they were originally assigned through randomization \citep{Brody2016}. In this method, patients who switch treatments are censored or simply excluded at the time of crossover. We refer to these two PP approaches as censor-at-switching (CAS) and exclude-at-switching (EAS) respectively.

The PP method is widely used as a sensitivity analysis to show the robustness of the ITT analysis. However, in contrast to ITT analysis, which makes complete use of the patients' information and guarantees comprehensive balance between the control group and the treatment group, PP methods may introduce selection bias due to censoring or excluding parts of the balanced data \citep{Brody2016}. The treatment effect estimated by CAS and EAS implicitly assumes that the {\revise crossed-over} patients are exchangeable with those that remained in the control group. 

For the patients in the control group, recall that $T$ is the time to event and $U$ is the time to crossover, where $U = \infty$ if $U>T$ (since crossover after the event cannot be observed).
Assuming that patients drop out of the clinical trial at time $V$, we have the event indicator $\delta=I(T\le U, T\le V)$. Unlike the traditional ITT approach, patients who switch to the experimental treatment are \emph{also} considered as censored under the PP method. 

Due to its reliance solely on data before the crossover point, CAS and EAS only provide estimates of the hazard function $\lambda_{1}(t)$ before crossover \eqref{lambda1}. 
 However, researchers are typically interested in estimating the overall hazard function $\widetilde{\lambda}_{B}(t)= P(T=t\mid T\ge t)$. Clearly, $\lambda_1(t) \ne \widetilde{\lambda}_{B}(t)$ unless $\lambda_{2}^{\bf x}(t\mid u)=\lambda_{1}(t)$. If this holds, then we have a Markov model where the hazard functions before and after the crossover are the same. Therefore, the CAS and EAS methods implicitly assume Markov crossover \eqref{Markov_assumption}.

\subsection{Treatment as Time-Dependent Variable}
\label{sec:TTDV}

Building upon the Cox proportional hazards (PH) model \citep{Cox1972}, we can introduce the treatment assignment as a covariate that changes over time \citep{White1997, Morden2011}. This allows for evaluation of the influence of the treatment that a patient actually undergoes. This model can be represented as a Cox PH model,
\begin{equation} \label{TTDV}
    \lambda(t\mid X(t))=\lambda_{BL}(t) \exp\{\beta X(t)\},
\end{equation}
where $\lambda_{BL}(t)$ represents the baseline hazard function, and we assume $X(t)=0$ when the patient is in the control group at time $t$ and $X(t)=1$ when the patient is in the experimental treatment group at time $t$. We refer to \eqref{TTDV} as a ``treatment as time-dependent variable'' (TTDV) model. 

For the patients in the control group, using the time-dependent covariate $I(t>u)$, we have the hazard function,
\begin{eqnarray} \label{TTDV2}
    &&P(T=t\mid T\ge t,U=u)=\lambda_{BL}(t)\exp\{\beta I(t>u)\}.
\end{eqnarray}
This implies that the hazard function {\revise from entry to event without crossover is}  $\lambda_1(t)=\lambda_{BL}(t)$ and the hazard function {\revise from crossover to event is} $\lambda_2^{\bf x}(t\mid u)=\lambda_{BL}(t)\exp(\beta)$. From \eqref{TTDV2}, we see that TTDV is a Markov model \eqref{Markov_assumption} since the hazard function after crossover does not depend on the crossover time $u$. The TTDV model \eqref{TTDV2} also implies that the hazard ratio $\mbox{HR}(t)$ in \eqref{treatmenteffect} is constant.

Similar to the previously mentioned PP method, the TTDV approach has the potential to disrupt the randomization assumption and may consequently introduce selection bias when switching is linked to the patient's prognosis \citep{White1999}.

\subsection{Rank Preserving Structural Failure Time Models}
 \label{sec:RPSFT}
 
\cite{James1991} proposed rank preserving structural failure time (RPSFT) models to estimate the true treatment effect under an accelerated failure time (AFT) structural model. The time at which an event is observed in a patient can be used to infer the counterfactual time at which the same event would have been observed if the {\revise crossed-over} patient had \emph{not} undergone any experimental treatment. These models are called ``rank preserving'' because they assume that if two patients $i$ and $j$ had the same treatment, with patient $i$ experiencing the event before patient $j$, the same order for the time-to-event would hold if both patients were given an alternative treatment \citep{James1991}.

For individuals that switched to the experimental treatment, let $T^*$ and $U$ be the event time and the crossover time respectively. Letting $T$ denote the counterfactual survival time (i.e. the survival time that the patient would have had if {\revise (s)he} had remained in control group), the RPSFT method posits the model
\begin{equation} \label{rpsft}
    T=U+e^{-\phi_0}(T^*-U),
\end{equation}
where $e^{\phi_0}$ is a so-called acceleration factor indicating the degree by which a patient's expected time to an event is extended due to the experimental treatment. If $e^{\phi_0} > 1$, this suggests a positive treatment effect, while the unusual $e^{\phi_0} < 1$  represents a negative treatment effect. In either case, the treatment effect is assumed to be constant. 

Similar to the TTDV method \eqref{TTDV2}, we can define a time-dependent covariate $I(t>U)$. Assume there is a constant treatment effect for patients who are switched to a different group from the one to which they were originally assigned. Then \eqref{rpsft} can be rewritten as 
\begin{equation} \label{rpsft2}
    T=\int_{0}^{T^*} \exp(\phi_{0} I(t>U)) dt.
\end{equation}
The formulation \eqref{rpsft2} also applies to patients who did not switch treatments. In that case, we can set $U=T^*$ for patients remaining in the control group or $U=0$ for the patients remaining in the treatment group respectively for the whole duration of the study. 

The RPSFT method is inherently a semi-Markov model \eqref{semi-Markov_assumption}. To better understand this, let us again split the hazard function after the crossover point into two parts,
\begin{eqnarray*}
\lambda_2^{\bf x}(t\mid u)=\left\{
\begin{array}{ll}
\lambda_2(t-u), &\quad\mbox{if patient remains in control group,}\\ \lambda_{2*}(t-u), &\quad\mbox{if patient switches treatment.} 
\end{array}
\right.
\end{eqnarray*}
Based on \eqref{rpsft}, it is equivalent to connect the hazard functions $\lambda_{2*}$ and $\lambda_{2}$ in \eqref{lambda2star_and_lambda2} via
\begin{eqnarray} \label{lambda2lambda2star}
    \lambda_{2*}(t)=e^{-\phi_{0}}\lambda_{2}(te^{-\phi_{0}}),
\end{eqnarray}
such that they have the same survival probability at time $t$ and $te^{-\phi_{0}}$, i.e.
\begin{equation*}
    S_{\lambda_{2*}}(t)=S_{\lambda_{2}}(te^{-\phi_{0}}) 
\end{equation*}
It should be noted that the model \eqref{lambda2lambda2star} is not identifiable if {\revise all the patients} in the control group cross over to the experimental treatment group. In this scenario, the hazard function $\lambda_{2}$ is not estimable since there is no available data for patients who {\revise continue} the control treatment. To make the model identifiable, one typically needs to introduce an additional assumption,
\begin{equation} \label{identifiability1}
    \widetilde{\lambda}_{A}(t)=e^{-\phi_0}\widetilde{\lambda}_{B}(te^{-\phi_0}).
\end{equation}
This assumption implies that the constant treatment effect is the same for all patients, regardless of when they receive it.

\subsection{{\revise Two-Stage AFT-Model Estimation Method}} \label{sec:TSAFT}

{\revise \cite{Nicholas2017,Nicholas2018} described a two-stage estimation (TSE) adjustment method to account for treatment crossover that happens after a specific disease-related time point, such as disease progression, which is known as a ``secondary baseline".  In the first stage of the TSE method, the focus is on estimating the impact of switching on survival after the secondary baseline. The second stage then uses this estimated effect to calculate counterfactual survival times for the crossed-over patients. \citep{Nicholas2017,Nicholas2018}.

Specifically, for individuals that switched to the experimental treatment, let $T^*$ and $U$ be the event time and the crossover time respectively. Let $T$ denote the counterfactual survival time. In the first stage, a standard parametric accelerated failure time model (AFT), such as a Weibull or generalized gamma AFT regression, is often used to estimate the treatment effect after the secondary baseline \citep{Nicholas2017}. Once the treatment effect associated with switching (denoted as $e^{\phi_2}$, with $\phi_2\ge 0$) is estimated, stage two of the TSE method uses the inverse of the estimated treatment effect to derive the counterfactual survival time $T$ as
\begin{equation} \label{tse}
    T=U+e^{-\widehat{\phi}_2}(T^*-U).
\end{equation}
Using a similar AFT model, a comparison between the treatment group and the control group will then be made based on the counterfactual survival times to derive the ``treatment effect" as if crossover had not been allowed. We denote this treatment effect as $e^{\phi_1}$, with $\phi_1\ge 0$. 

The TSE method is a semi-Markov model for the same reasons as the RPSFT method. However, the main difference between TSE and RPSFT is that RPSFT imposes the assumption that $\phi_1=\phi_2$. The constant treatment effect assumption of RPSFT assumes that the treatment effect is consistent across both the experimental groups and the crossed-over patients. Relaxing this assumption in the TSE method broadens its applicability, especially in oncology trials with post-progression switching, since progressive disease could impact the capacity to benefit from treatment. However, by making the constant effect assumption and using $g$-estimation tailored for an RCT context, RPSFT avoids the need to impose the ``no unmeasured confounding'' assumption \citep{James1991, Mark1993AMF}. On the other hand, this assumption is required for the TSE method. In this context, no unmeasured confounding means that crossover at any time point is independent of the counterfactual survival time $T$.

In the scenario with $100\%$ crossover, the treatment effect associated with switching $e^{\phi_2}$ is not identifiable, and therefore neither is $e^{\phi_1}$. In this case, the TSE method does not work. Some additional assumption like $\phi_1=\phi_2$ is often imposed, which essentially makes the TSE method the same as the RPSFT method. Some other assumption, such as $\phi_1=d\phi_2$ with a pre-specified $d>0$, would also remove the non-identifiability issue of the TSE method. However, none of these assumptions is testable based on data with $100\%$ crossover. 

Note that the treatment effects $e^{\phi_1}$ and $e^{\phi_2}$ in the TSE method can also be estimated with Cox PH models, in which case, the counterfactual event time needs to be modified accordingly.  However, we follow the original proposal by \cite{Latimer2017} and use the two-stage parametric AFT model for the estimation. We call this variant of the TSE method the two-stage AFT-model estimation (TSAFT) method to be more precise.
}

\subsection{Inverse-Probability-of-Censoring Weighting}
\label{sec:IPCW}
The RPSFT {\revise and TSAFT methods incorporate all available patient data and adjust the survival time} to account for what might have occurred if the patients who switched treatments had stayed in the control group. On the other hand, the inverse-probability-of-censoring weighting (IPCW) approach \citep{curtis2007using} focuses on the survival time \emph{before} the crossover by marking patients as censored at the point of treatment switching in the analysis.
As previously mentioned for the PP method (Section \ref{sec:PP}), this introduces bias because patients whose event time is censored tend to have systematic differences in prognosis compared to those whose who do not switch treatments. 

To correct this bias, patients in the control group who did \emph{not} switch to the experimental treatment can be assigned weights to account for the absence of data. In the IPCW method, the bias caused by informative crossover is adjusted by assigning each patient a weight that is the reciprocal of their estimated probability of not experiencing censoring at a specific time point \citep{robins2000correcting}. IPCW estimates the likelihood of patients switching treatments based on their individual baseline characteristics and time-dependent covariates. This estimation is often done using a logistic regression \citep{curtis2007using, robins2000correcting}.

The IPCW approach assumes no unmeasured confounders at the given time of crossover, making the censoring noninformative after inverse-probability weighting \citep{curtis2007using, robins2000correcting}. In essence, this assumption implies that if one has adequately considered and controlled for all the relevant covariates that could influence both the treatment assignment and the outcome, the results obtained from this analysis will provide a less biased estimate of the true treatment effect. However, if there \emph{are} unmeasured or unaccounted-for factors that confound the relationship, the results may be biased. Ensuring that this assumption is reasonably met is crucial to making valid inferences and drawing accurate conclusions from observational data. In practice, researchers often use techniques like propensity score weighting or matching to address potential confounding covariates \citep{Austin2011}.

Recall that $T$ is the time to event, $U$ is the crossover time, and let $V$ be the censoring time in the data. The event indicator in the IPCW approach is $\delta=I(T\le U, T\le V)$. Let $Y = \min(T, U, V)$. The IPCW approach further makes the assumption that $T$ and $U$ are conditionally independent given some (potentially time-dependent) covariates $X$. If this assumption holds, then 
\begin{eqnarray*}
    P(Y=t,\delta=1\mid X)=P(T=t\mid X)\times P(U\wedge V\ge t\mid X).
\end{eqnarray*}
If we can reliably estimate the {\revise survival function for censoring} $W(t,x)=P(U\wedge V\ge t\mid X=x)$ as $\widehat{W}(t, x)$, then the overall hazard function $\widetilde{\lambda}_B(t\mid X=x)=P(T=t\mid T\ge t, X)$ can be estimated via inverse-weighting of the observed event times as 
\begin{eqnarray*}
    \widehat{\lambda}_B(t\mid X=x)=\frac{\sum_{i=1}^n I(Y_i=t,\delta_i=1, X_i=x)/\widehat{W}(t,X_i)}{\sum_{i=1}^n I(Y_i\ge t, X_i=x)/\widehat{W}(t,X_i)}.
\end{eqnarray*}
However, if the conditional independence assumption does \emph{not} hold, then this method estimates
\begin{equation*}
    \lambda_1(t\mid X) = P(T=t\mid T\ge t,U\ge t, X), \hspace{0.5cm}t>0,
\end{equation*}
which is the hazard function {\revise without} crossover. This hazard function in general is not equal to the overall hazard function, {\revise i.e.,}
\begin{equation*}
    P(T=t\mid T\ge t,U\ge t, X)\ne P(T=t\mid T\ge t, X), \hspace{0.5cm}t>0.
\end{equation*}


\subsection{New Method: Bayesian Imputed Multiplicative Method} \label{sec:new_method}

As discussed previously, the TSM framework not only unifies existing approaches like ITT, CAS, EAS, TTDV, RPSFT, {\revise TSAFT,} and IPCW (see Table \ref{tab:summary}), but it \emph{also} facilitates new methods to adjust for confounding from treatment crossover. In some treatment crossover scenarios, a researcher may want to invent a new method for treatment effect estimation tailored to a specific scenario or set of assumptions. In this section, we propose a new model under the TSM framework that is particularly well-suited when the crossover is informative and linked to the occurrence of a disease-related event like disease progression. 

As before, let $T^*$ and $U$ be the event time and the crossover time respectively for individuals that switched to the experimental treatment. Let $T$ denote the counterfactual survival time. We assume semi-Markov crossover \eqref{semi-Markov_assumption}, i.e., 
\begin{eqnarray*}
\lambda_2^{\bf x}(t\mid u)=\left\{
\begin{array}{ll}
\lambda_2(t-u), &\quad\mbox{if patient remains in control group,}\\ \lambda_{2*}(t-u), &\quad\mbox{if patient switches to treatment.} 
\end{array}
\right.
\end{eqnarray*}
If we assume an AFT model $
    \lambda_{2*}(t)=e^{-\phi_{0}}\lambda_{2}(te^{-\phi_{0}})$,
then we have $T=U+e^{-\phi_0}(T^*-U)$, which is the RPSFT method \eqref{rpsft}. Alternatively, we can assume a \textit{multiplicative hazard} model, $\lambda_{2*}(t)=e^{\beta}\lambda_{2}(t)$. Then the counterfactual survival time $T$ can be expressed as
\begin{equation} \label{newmethod}
    T=U+S^{-1}_{2}[\{S_{2*}(T^*-U)\}],
\end{equation}
where $S_{2*}$ is the survival function based on the hazard function $\lambda_{2*}$, and $S^{-1}_{2}$ is the inverse function of the survival function $S_{2}$ based on the hazard function $\lambda_2$. It should be stressed that in general, the treatment effect after crossover $\lambda_{2*}(t) / \lambda_{2}(t)$ is not the same the overall treatment effect $\widetilde{\lambda}_A(t) / \widetilde{\lambda}_B(t)$. 

Let $\pi_2$ represent the percentage of patients who switched to the experimental treatment. 
As long as $\pi_2 < 1$, the model \eqref{newmethod} is identifiable under the semi-Markov assumption. When $\pi_2=1$, the hazard function $\lambda_2$ cannot be directly estimated since there is no data to estimate $\lambda_2(t)$. To make the model identifiable when $\pi_2 = 1$, we need to impose an additional assumption. Specifically, we assume that the treatment effect after crossover is the same as the overall treatment effect between the treatment group and the control group when crossover is not allowed, i.e.,
\begin{align} \label{additional_assumption}
        \lambda_{2*}(t)=e^{\beta}\lambda_{2}(t) \quad \mbox{and} \quad   \widetilde{\lambda}_{A}(t)=e^{\beta} \widetilde{\lambda}_{B}(t), \hspace{.4cm} \text{ if } \pi_2 = 1.
\end{align}
In other words, if there is 100\% crossover, e.g. in the CENTAUR trial \citep{Paganoni2022} described in Section \ref{sec:intro}, then we need to find an estimate of $e^{\beta}$ that meets this assumption \eqref{additional_assumption}. In practice, it may not be easy to verify this assumption. Thus, achieving accurate estimates of the overall treatment effect \eqref{treatmenteffect} might be very difficult under 100\% crossover.

To implement our method, we model the hazard functions as piecewise constant functions taking the form,
\begin{equation} \label{piecewiseconstant}
    \lambda(t) = \sum_{j=1}^{J} \xi_{j} \mathbb{I} \{ t \in ( s_{j-1}, s_j] \},
\end{equation}
where $\xi_{j}$ is the hazard in the time interval $(s_{j-1},s_{j}]$ and $0 = s_0 < s_1 < s_2 < \cdots < s_J < \infty$, where $s_J$ is larger than the largest observed time in the study. Note that the cut points $s_j$'s for different hazard functions $\lambda_1(t)$, $\lambda_2(t)$, $\lambda_{2*}(t)$ and $\lambda_3(t)$ are the same, but the hazard constants are different. {\revise Furthermore, the piecewise exponential model can approximate any arbitrary hazard function (continuous or noncontinuous) as long as the pieces are small enough.}

To estimate these hazard functions, we adopt a Bayesian approach and endow the hazards $\xi_j$'s in \eqref{piecewiseconstant} with weakly informative independent Gamma priors \citep{IbrahimSpringer2001}. {\revise We found that the results from using weakly informative Gamma priors were practically indistinguishable from those using noninformative improper priors $\lambda_j \propto 1, j = 1, \ldots, J$. Moreover, the proper Gamma prior tended to be a more numerically stable choice.} We call our approach the Bayesian imputed multiplicative method (BIMM). Assuming no covariates, the BIMM method estimates the log hazard ratio $\beta$ and corresponding variance as follows.
\vspace{.5cm}

\noindent \textbf{When $\pi_2 < 1$:}
\begin{enumerate}[i.]
\item Use Markov chain Monte Carlo (MCMC) to estimate $\lambda_1(t),\lambda_2(t), \lambda_{2*}(t), \lambda_3(t)$ as in \eqref{lambda1}-\eqref{lambda3} using \eqref{piecewiseconstant} with Gamma priors on the step heights $\xi_j$'s. 
\item For each MCMC sample $k=1,\ldots,K$:
\begin{enumerate}[a.] 
\item Compute the counterfactual survival time using \eqref{newmethod} for the control patients who switch to treatment. 
\item Fit the Cox PH model comparing the treatment group data with the adjusted control group data. The Cox PH model will give the estimate of log hazard ratio and the corresponding model-based variance, denoted by $\beta_k$ and $v_k$.
\end{enumerate}
\item Summarize the $K$ fitted Cox PH models so that the point estimate of $\beta$ is the mean of $\beta_k$, $k=1,\ldots, K$, and the variance estimate is the sum of the mean of the $v_k$'s and the sample variance of the $\beta_{k}$'s. 
\end{enumerate} 


{\revise
\noindent \textbf{When $\pi_2 = 1$:}
\begin{enumerate}[i.]
\item Use MCMC to estimate $\lambda_1(t),\lambda_{2*}(t), \lambda_3(t)$ as in \eqref{lambda1}-\eqref{lambda3} using \eqref{piecewiseconstant} with Gamma priors on the step heights $\xi_j$'s. 
\item For each MCMC sample $k = 1, \ldots, K$:
\begin{enumerate}[a.]
\item Set $m=0$ and initialize $\beta_k^{(m)}$. Calculate $\lambda_{2}(t)$ as $\lambda_{2}(t)=\lambda_{2*}(t)/e^{\beta_k^{(m)}}$
\item Compute the counterfactual survival time using \eqref{newmethod} for the control patients who switch to treatment.
\item Fit the Cox PH model comparing the treatment group data with the adjusted control group data. The Cox PH model will give the estimate of the log hazard ratio $\beta_k^{(m+1)}$ and the corresponding model-based variance $v_k^{(m+1)}$. 
\item Repeat steps (a)-(c) until the sequence $\{\beta_k^{(m)}, m \ge 0\}$ converges. In practice, we use the convergence criterion $|\beta_k^{(m+1)} - \beta_k^{(m)} | < 10^{-6}$.
\item Once the model has converged, denote the point estimate and variance estimate as $\beta_k$ and $v_k$ respectively. 
\end{enumerate}
\item Summarize the $K$ fitted Cox PH models so that the point estimate of $\beta$ is the mean of $\beta_k$, $k = 1, \ldots, K$, and the variance estimate is the sum of the mean of the $v_k$'s and the sample variance of the $\beta_k$'s.
\end{enumerate} 
}

\section{Simulations} \label{Sec:sims}

To illustrate our TSM framework, we conducted simulation experiments in scenarios where treatment crossover was allowed. The first experiment mimics a clinical trial design, where the hazard functions were specified in relatively simple forms. In the second experiment, we re-engineered individual patient data from a real clinical trial and then used piecewise exponential models to estimate the corresponding hazard rates. We then simulated data based on these hazard rates. Each experiment was repeated for 2000 replications using the seven methods described in Section \ref{sec:TSM_methods} and reported in Table \ref{tab:summary}: ITT (Section \ref{sec:ITT_method}), CAS (Section \ref{sec:PP}), EAS (Section \ref{sec:PP}), TTDV (Section \ref{sec:TTDV}), RPSFT (Section \ref{sec:RPSFT}), {\revise TSAFT (Section \ref{sec:TSAFT})}, IPCW (Section \ref{sec:IPCW}), and BIMM (Section \ref{sec:new_method}). For each of these methods, we estimated both the treatment effect and the 95\% confidence interval (CI) for the treatment effect. {\revise Codes for implementing all of these methods are available at \url{https://github.com/JasonZhao111/TSM}.} 

For ITT, CAS, EAS, TTDV, and IPCW, we used a Cox PH model that was fit with the \textsf{R} package \texttt{survival} (version 3.5-7, available on the Comprehensive \textsf{R} Archive Network (CRAN)). For RPSFT, we used the \textsf{R} package \texttt{rpsftm} (version 1.2.8 on CRAN) to estimate $\phi_{0}$ in \eqref{rpsft}. We then used a Cox PH model to calculate the log hazard ratio $\widehat{\beta}$ based on the observed survival times in the control group and the counterfactual survival times in the treatment group adjusted by plugging in the estimated acceleration factor $\hat{\phi}_{0}$ in \eqref{rpsft}. To estimate the RPSFT standard error for $\widehat{\beta}$, we followed \citet{BENNETT2018} and \citet{Robin1991} and used $\text{se}(\widehat{\beta}) = | \widehat{\beta} | / \sqrt{\chi_{itt}^2}$, where $\chi^2_{itt}$ represents the chi-squared test statistic from the log-rank test for ITT analysis comparing the original data between control group and treatment group. {\revise For TSAFT, we used the \texttt{survival} package to estimate $\phi_2$ in \eqref{tse} using a parametric AFT model with a Weibull distribution. We then plugged in the estimated $\widehat{\phi_2}$ into \eqref{tse} to obtain the counterfactual survival times in the treatment group and used a Cox PH model to calculate the log hazard ratio $\widehat{\beta}$. To estimate the standard error for $\widehat{\beta}$ under TSAFT, we used the nonparametric bootstrap with 200 replicates.}

For IPCW, the probability of \emph{not} crossing over was estimated by a logistic regression in the control group, as suggested by \cite{Ishak2014}.  It is important to note that IPCW method is numerically unstable when the proportion of switching is very high (e.g. 100\% crossover). In this case, the inverse probability of not switching may be extremely large or nonestimable. Therefore, we set an upper bound on the inverse probability to be 10 in order to avoid extremely large weights.

Finally, the BIMM method was implemented as described in Section \ref{sec:new_method} using \texttt{Stan} \citep{carpenter2017stan} interfaced with \textsf{R} through the package \texttt{Rstan} (version 2.32.5 on CRAN). We specified the priors on the hazards $\xi_j$'s in \eqref{piecewiseconstant} to be {\revise weakly informative} Gamma(1{,}2) {\revise priors}. We ran eight MCMC chains of 2000 iterations each and discarded the first 500 iterations of each chain as burnin, leaving us with a total of 12,000 MCMC samples with which to estimate the hazard functions. In Experiment 1, each of the hazards $\xi_j$'s corresponded to the time interval $(j-1,j]$, $j=1,...,5$. In Experiment 2 based on reverse-engineered data from a real clinical trial, we evenly divided the time intervals into four pieces from entry to the maximum observed survival time. 



\subsection{Experiment 1: Simulated Trials} \label{subsec:ex1}
\subsubsection{Simulation Settings} \label{subsubsec:ex1_settings}

In our first experiment, data were generated to mimic a  clinical trial design in Oncology. We simulated data on $n=400$ total patients, with 200 patients initially belonging to the control group and 200 patients belonging to the experimental treatment group at randomization. 

We designed a survival trial based on the three-state model. All the hazard rates were yearly hazard rates with three pieces corresponding to years 0-1, 1-2 and 2+. Let $\lambda(\cdot)=c(a_1,a_2,a_3)$ denote hazard rates for years 0-1, 1-2 and 2+ as $a_1$, $a_2$ and $a_3$ respectively. For the control patients, we assumed that the crossover happens at the time of disease progression. The crossover is assumed to be semi-Markov, i.e. $\lambda_2^{\bf x}(t\mid u)=\lambda_2(t-u)$. We specified the hazard functions $\lambda_1(t)$ and $\lambda_3(t)$ as $\lambda_{1}(\cdot)=c(0.2,0.2,0.25)$ and $\lambda_{3}(\cdot)=c(0.4,0.4,0.4)$ respectively. For the control patients who switched to the experimental treatment, their hazard rate after crossover was $\lambda_{2*}(\cdot)=0.8\times\lambda_1(\cdot)$. For the control patients who (potentially) remained in the control group, their hazard rate was $\lambda_{2}(\cdot)=1.5\times \lambda_1(\cdot)$. As such, patients who switched to the experimental treatment had lower hazard rates than those who remained in the control after disease progression. For the patients allocated to the treatment arm, no crossover was allowed, so we assumed a piecewise exponential model with hazard rates $\widetilde{\lambda}_A(\cdot)=c(0.12,0.12,0.15)$ for the survival time. In other words, the HR between $\widetilde{\lambda}_A(\cdot)$ and $\lambda_1(\cdot)$ was $0.6$. 

We assumed that the yearly censoring rate was {\revise $0.02$}. The study had a {\revise one-year} recruitment period with a uniform accrual rate to recruit $400$ patients, and the study would be read out at {\revise 6} years after the randomization of the first patient. {\revise  In Appendix \ref{App:A}, we present more simulation results under lower and higher censoring rates.} 



Let $\pi_2$ denote {\revise the proportion of control patients that cross over to the experimental treatment}; then $1-\pi_2$ denotes {\revise the proportion of control patients that remain in the control group}. We {\revise considered four different crossover rates $\pi_2 \in \{0.25,0.5,0.75,1\}$} in our simulations. {\revise For the different crossover proportions, we obtained the following HRs, statistical powers, and censoring proportions. For \(\pi_2 = 0.25\), the HR was 0.544, with power of 99.8\% and a censoring proportion of 38.4\%. For \(\pi_2 = 0.5\), the HR increased to 0.592, with power of 98.3\% and a censoring proportion of 40.0\%. For \(\pi_2 = 0.75\), the HR further increased to 0.643, accompanied by a decrease in power to 92.2\% and an increase in censoring proportion to 41.6\%. Finally, for \(\pi_2 = 1\), the HR reached 0.699, with power of 77.0\% and the censoring proportion 43.2\%. }It is important to note that the hazard ratio $\mbox{HR}(t)$ is not a constant over time. Rather, these HRs were obtained by fitting the data as if the proportional hazards assumption holds. This approach reflects the most common way of reporting the results in clinical study reports. Therefore, we used the same approach in this experiment. 

For comparison, if control patients were \emph{not} allowed to switch to the treatment group after disease progression, the HR would be {\revise $0.5$}. This is the \emph{true} treatment effect when crossover is \emph{not} allowed. The power in this case would be {\revise $99.9\%$}. Apparently, without adjusting for the crossover in the control group, we would have a biased estimate of the true treatment effect. All the results in the preceding and current paragraphs were obtained using the \textsf{R} package \texttt{PWEALL} (version 1.4.0 on CRAN). 

{\revise Finally, it should be stressed that our data were all simulated as \emph{non}-proportional hazards. Therefore, the data generating mechanism was never exactly the same as any of the models that we fit, allowing for a fair comparison between the different TSM methods.}

\subsubsection{Simulation Results}


\begin{table}[t!]
\centering
\captionof{table}{Simulation results for Experiment 1 based on 2000 replications. The table displays the average Bias, SE, and MSE from all Monte Carlo replicates, while the ECP is the percentage of 95\% CIs that contained the true HR.}
\begin{tabular}{l c cccc c cccc}  
\toprule
 & & \multicolumn{4}{c}{\textbf{25\% crossover}} & & \multicolumn{4}{c}{\textbf{50\% crossover}} \\
\cmidrule(lr){3-6} \cmidrule(lr){8-11}
\textbf{Method} & & Bias & SE & MSE & ECP (\%) & & Bias & SE & MSE & ECP (\%) \\
\midrule
ITT & & 0.047 & 0.071 & 0.012 & 90.5 & & 0.097 & 0.078 & 0.021 & 75.1 \\
CAS & & -0.034 & 0.060 & 0.008 & 92.5 & & -0.073 & 0.056 & 0.011 & 77.7 \\
EAS & & -0.024 & 0.063 & 0.009 & 94.3 & & -0.060 & 0.061 & 0.011 & 84.4 \\
TTDV & & 0.025 & 0.068 & 0.010 & 93.6 & & 0.053 & 0.073 & 0.013 & 89.6 \\
RPSFT & & 0.006 & 0.082 & 0.012 & 96.2 & & 0.013 & 0.096 & 0.015 & 97.5 \\
TSAFT & & 0.005 & 0.068 & 0.009 & 94.4 & & 0.012 & 0.073 & 0.010 & 93.7 \\
IPCW & & -0.029 & 0.062 & 0.008 & 93.6 & & -0.072 & 0.059 & 0.012 & 79.5 \\
BIMM & & 0.016 & 0.069 & 0.010 & 94.8 &  & 0.020 & 0.073 & 0.011 & 94.6 \\
\end{tabular} \\[.1in]

\begin{tabular}{l c cccc c cccc}  
 & & \multicolumn{4}{c}{\textbf{75\% crossover}} & & \multicolumn{4}{c}{\textbf{100\% crossover}} \\
\cmidrule(lr){3-6} \cmidrule(lr){8-11}
\textbf{Method} & & Bias & SE & MSE & ECP (\%) & & Bias & SE & MSE & ECP (\%) \\
\midrule
ITT & & 0.145 & 0.085 & 0.035 & 52.6 & &0.204 & 0.094 & 0.060 & 29.6 \\
CAS & & -0.118 & 0.052 & 0.019 & 46.9 & & -0.168 & 0.047 & 0.033 & 15.0 \\
EAS & & -0.119 & 0.056 & 0.020 & 52.2 & & -0.217 & 0.046 & 0.052 & 6.25 \\
TTDV & & 0.088 & 0.083 & 0.021 & 80.9 & & 0.148 & 0.102 & 0.043 & 65.7 \\
RPSFT & & 0.019 & 0.114 & 0.022 & 97.4 & & 0.033 & 0.142 & 0.037 & 97.6 \\
TSAFT & & 0.017 & 0.084 & 0.014 & 93.7 & & -0.030 & 0.131 & 0.035 & 93.5 \\
IPCW & & -0.139 & 0.052 & 0.025 & 35.7 & & -0.257 & 0.038 & 0.069 & 0.05 \\
BIMM & & 0.017 & 0.079 & 0.013 & 94.6 &  & 0.028 & 0.077 & 0.022 & 79.7 \\
\bottomrule
\end{tabular} \label{tab:crossover_results}
\end{table}

In each of the 2000 replicates, we recorded the estimated treatment effect  $\widehat{\text{HR}}$ and the nominal 95\% confidence interval for the HR using the methods in Table \ref{tab:summary}. Table \ref{tab:crossover_results} reports the average bias ($\text{Bias} = \widehat{\text{HR}} - \text{HR}_{\text{true}}$), empirical standard error (SE), average mean-squared error ($\text{MSE} = \text{Bias}^2 + \text{SE}^2$), and empirical coverage probability (ECP) for each of the seven methods. The MSE balances the trade-off between bias and variance. The ECP is the proportion of 95\% CIs that contained the true HR.

\begin{figure} 
\begin{subfigure}[h]{0.5\linewidth}
\includegraphics[width=\linewidth]{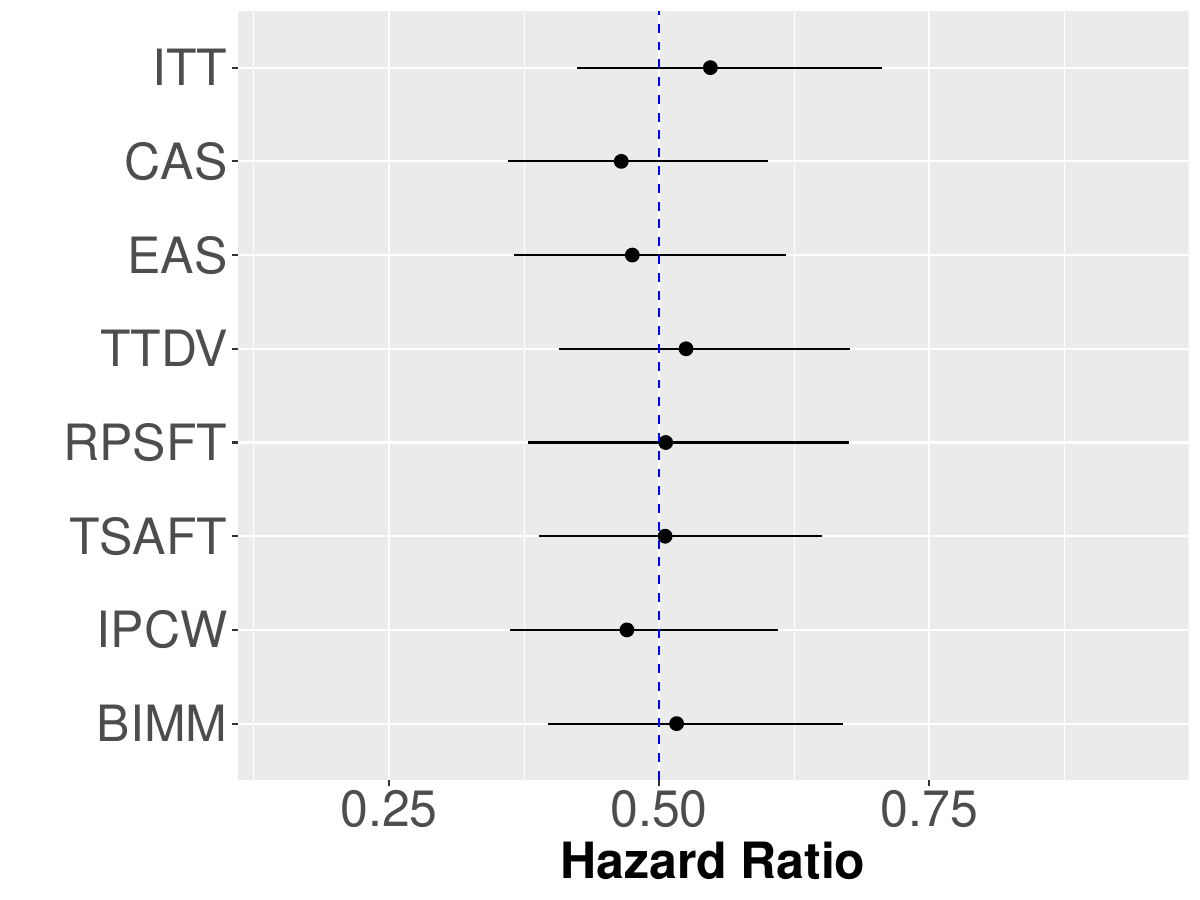}
\caption{25\% crossover after progression}
\end{subfigure}
\hfill
\begin{subfigure}[h]{0.5\linewidth}
\includegraphics[width=\linewidth]{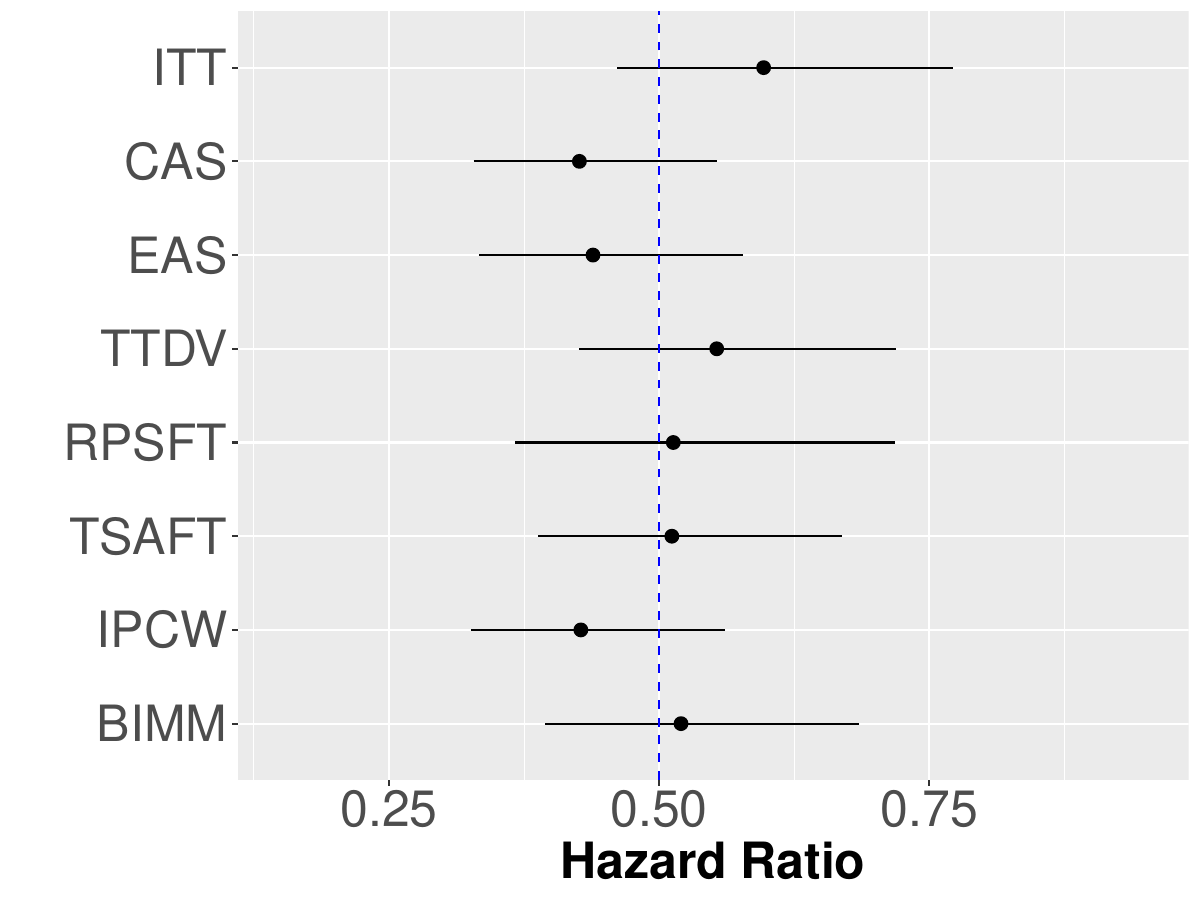}
\caption{50\% crossover after progression}
\end{subfigure}%
\\[.25in]
\begin{subfigure}[h]{0.5\linewidth}
\includegraphics[width=\linewidth]{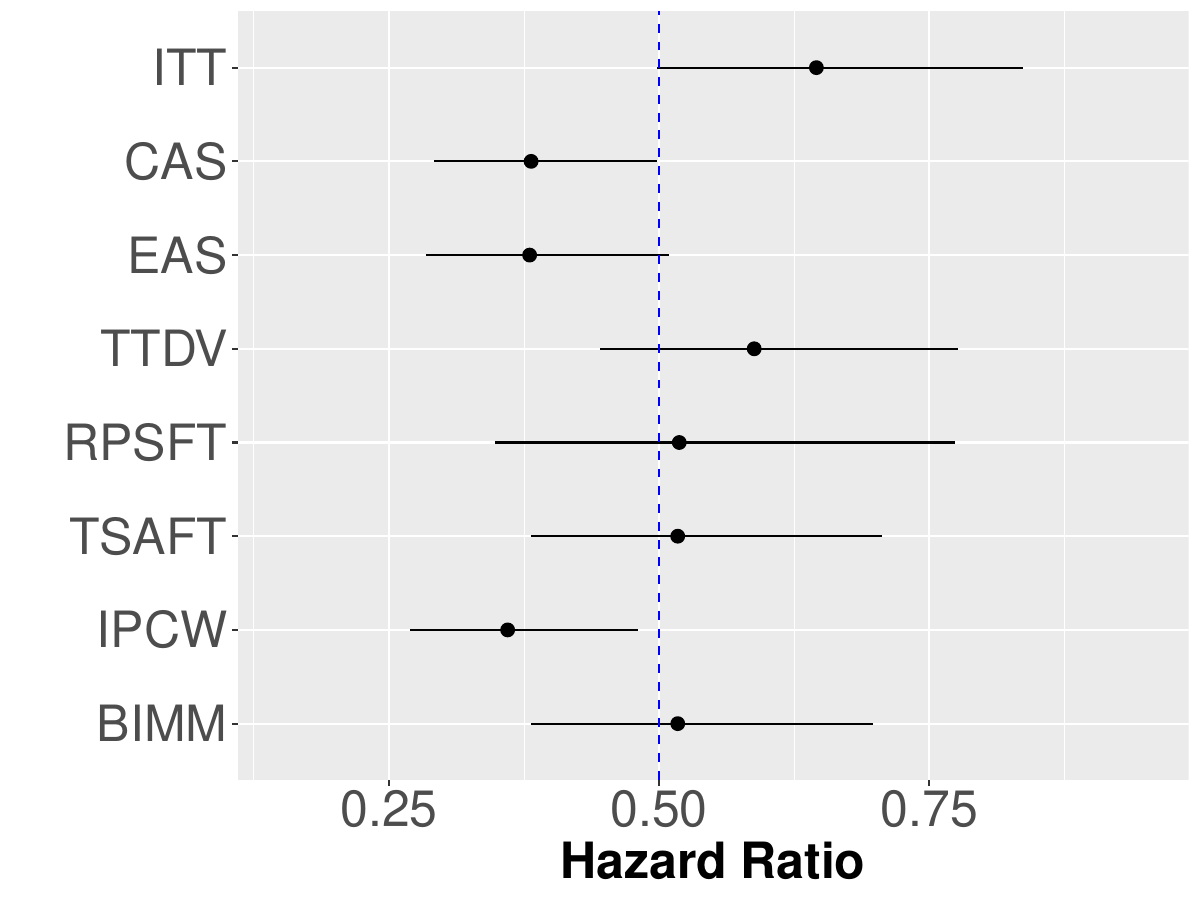}
\caption{75\% crossover after progression}
\end{subfigure}
\hfill
\begin{subfigure}[h]{0.5\linewidth}
\includegraphics[width=\linewidth]{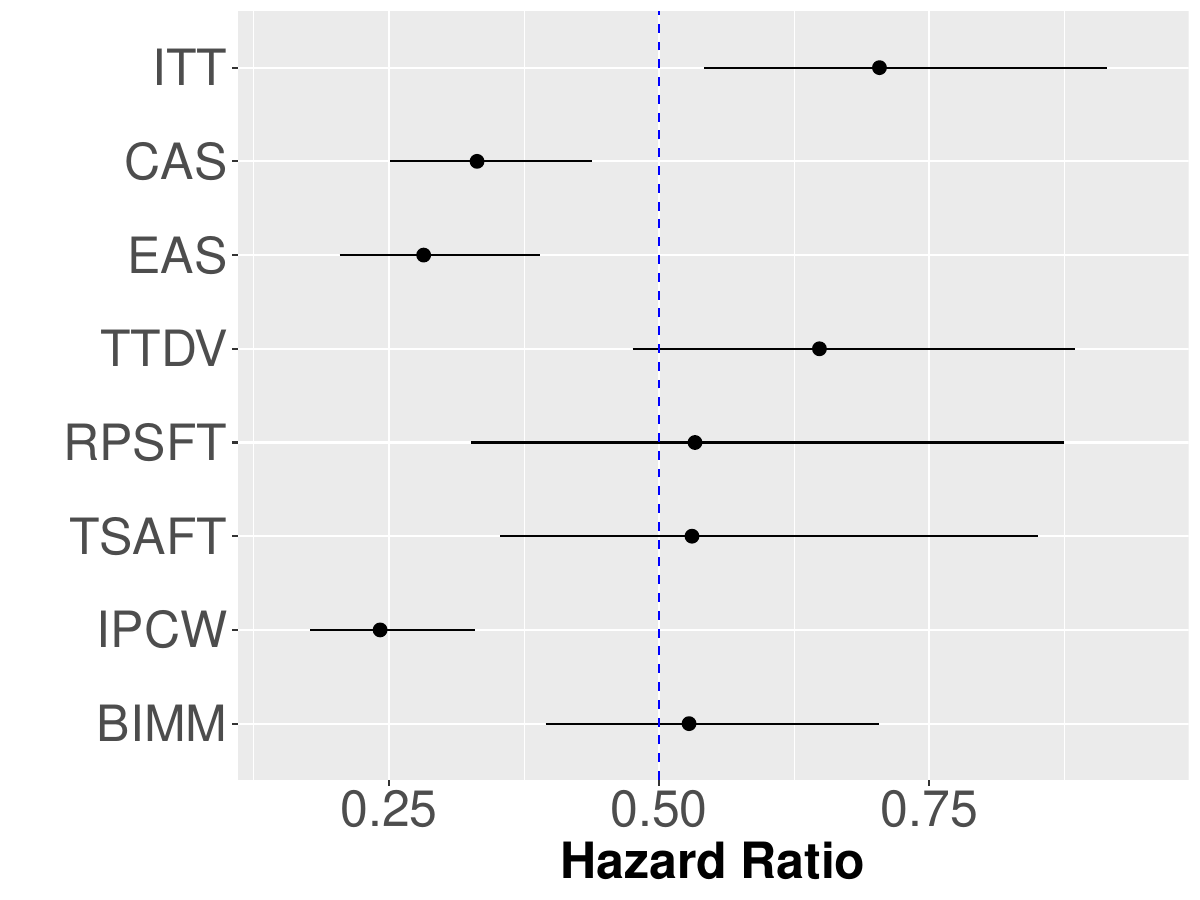}
\caption{100\% crossover after progression}
\end{subfigure}%
\caption{Simulation results in Experiment 1 averaged across 2000 replicates. {\revise The dotted blue line is the true treatment effect ($\text{HR}_{\text{true}}=0.5$).}} \label{fig:simulation_results}
\end{figure}

Table \ref{tab:crossover_results} shows that {\revise under 25\% crossover, all methods performed similarly in terms of average MSE, but TSAFT had the lowest average bias, followed by RPSFT and BIMM.  Under 50\% crossover, TSAFT also had the lowest average bias and lowest average MSE. When $\pi_2=0.50$, the average bias was also much lower and the ECP was much higher for the semi-Markov methods (RPSFT, TSAFT, and BIMM) than for the other methods. Once the crossover rate increased to 75\% and 100\%, however}, our new BIMM method had the lowest average bias \emph{and} the lowest average MSE. RPSFT had the highest ECP {\revise in all the crossover scenarios}. However, RPSFT had the highest average SE of all methods {\revise regardless of $\pi_2$}, and as depicted in Figure \ref{fig:simulation_results}, RPSFT's 95\% CIs were generally more conservative than those of the other methods. BIMM also had the second highest ECP with narrower CIs than RPSFT {\revise under 25\%, 50\%, and 75\% crossover. In these scenarios, BIMM achieved an ECP very close to the nominal level. Under 100\% crossover, TSAFT and RPSFT had the highest ECP, but their SEs were considerably larger (and therefore 95\% CIs were also considerably wider) on average than those under BIMM.} 

{\revise Table \ref{tab:crossover_results} shows that the performance of ITT, CAS, EAS, TTDV, and IPCW tended to degrade considerably as the crossover rate increased, with the average bias and average MSE sharply increasing and the ECP sharply decreasing as the crossover proportion $\pi_2$ increased. While all methods achieved fairly good coverage (or ECP) under 25\% crossover, this was not the case for 50\%, 75\% or 100\% crossover. ITT, CAS, EAS, TTDV, and IPCW had much higher average bias and MSE and much lower ECP than RPSFT, TSAFT, or IPCW in the cases of 75\% and 100\% crossover. On the other hand, RPSFT, TSAFT, and BIMM tended to be fairly robust to the amount of crossover, with a much smaller increase in mean bias and MSE under 100\% crossover.}

Figure \ref{fig:simulation_results} plots the mean estimated treatment effect and the mean 95\% confidence intervals (CIs) from 2000 replications. The mean 95\% CIs were calculated by taking the average of the left and right endpoints of the 95\% CIs from all 2000 replications. The true HR is depicted as a dashed blue vertical line. {\revise Figure \ref{fig:simulation_results} shows that under all scenarios}, our proposed BIMM method (Section \ref{sec:new_method}) provided improved accuracy in estimating the true HR. {\revise RPSFT and TSAFT also provided comparable improvement in the mean accuracy.} 

Under 100\% crossover, the bias of the treatment effect estimation depends on the validity of the assumptions imposed to ensure identifiability. We need to emphasize the significant challenge in verifying these assumptions because of the lack of data on patients remaining in the control group after progression. By incorporating an assumption of constant treatment effect \eqref{additional_assumption} before and after progression, the {\revise bottom right} plot in Figure \ref{fig:simulation_results} demonstrates that {\revise under 100\% crossover, RPSFT, TSAFT, and BIMM all} gave mean estimates that were closest to the true HR. However, the 95\% CIs for {\revise TSAFT and} RPSFT were quite wide {\revise in this setting. Meanwhile,} BIMM provided both greater accuracy and less variability in treatment effect estimation. 

{\revise Overall, our results suggest that in our simulation study, the methods which assumed semi-Markov crossover \eqref{semi-Markov_assumption} and which used the data after crossover (i.e. RPSFT, TSAFT, and BIMM) performed the best on average and were the most robust under different amounts of treatment crossover.}

 \subsection{Experiment 2: An Experiment Based on Data From a Clinical Trial} \label{Sec:LENVIMA}

On August 2014, researchers submitted a new drug application (NDA) seeking approval of lenvatinib (LENVIMA) using data from Study E7080-G000-303 (SELECT) \citep{SELECT2019}. LENVIMA is an anti-cancer medication for the treatment of certain types of thyroid cancer. In SELECT, a total of 392 eligible patients were randomly assigned in a 2:1 ratio to the LENVIMA arm and the placebo arm. They were either given LENVIMA at a dose of 24 mg through continuous once-daily oral administration or a matching placebo.
Patients received the blinded study drug once daily until any of the following events occurred: confirmed disease progression or withdrawal of consent. Patients in the placebo group, upon having their disease progression confirmed, had the option to request entry into the open-label LENVIMA treatment phase and receive experimental treatment \citep{SELECT2019}. As a result, it is a $100\%$ crossover for the placebo group. After the double-blind phase, patients who received LENVIMA and had not encountered disease progression had the option to request ongoing open-label LENVIMA treatment at the same dosage, as determined by the clinical judgment of the investigator \citep{SELECT2019}. 

\begin{figure}
 \begin{subfigure}{0.49\textwidth}
     \includegraphics[width=\textwidth]{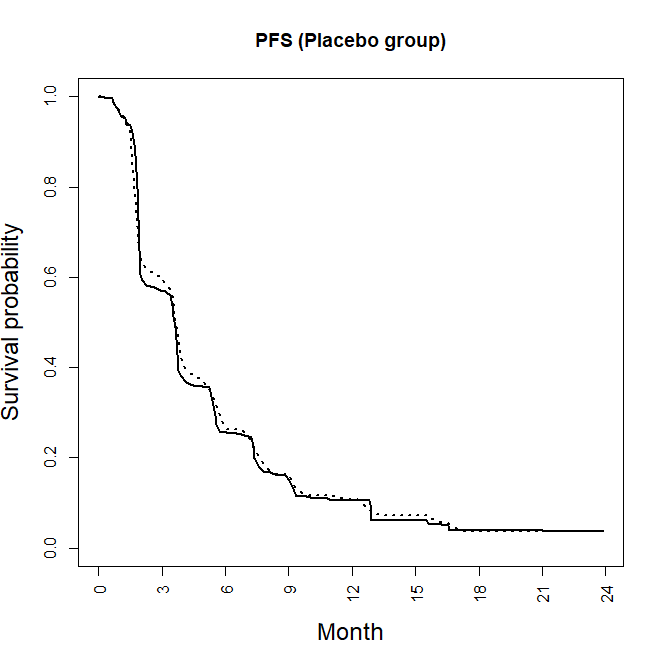}
 \end{subfigure}
 \hfill
 \begin{subfigure}{0.49\textwidth}
     \includegraphics[width=\textwidth]{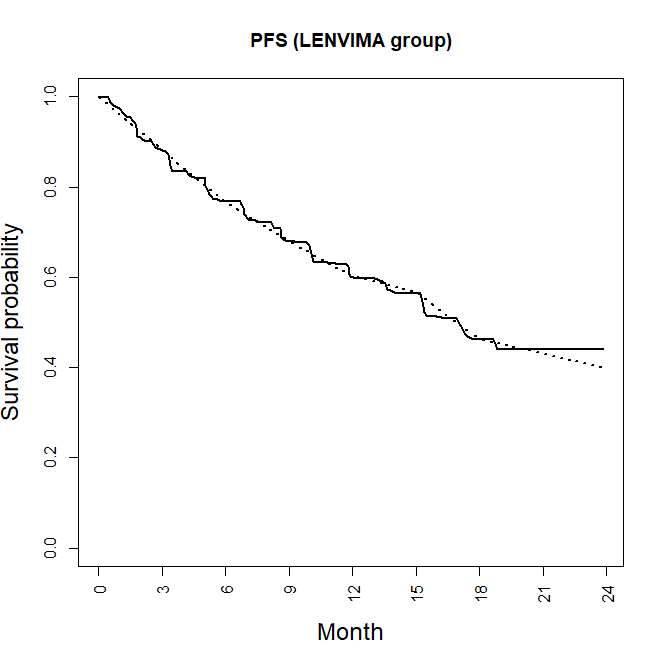}
 \end{subfigure}
 
 \medskip
 \begin{subfigure}{0.49\textwidth}
     \includegraphics[width=\textwidth]{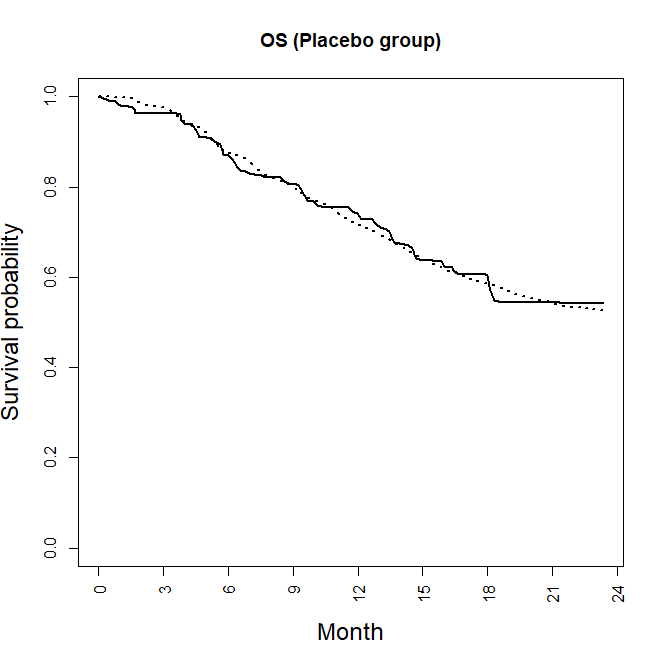}
 \end{subfigure}
 \hfill
 \begin{subfigure}{0.49\textwidth}
     \includegraphics[width=\textwidth]{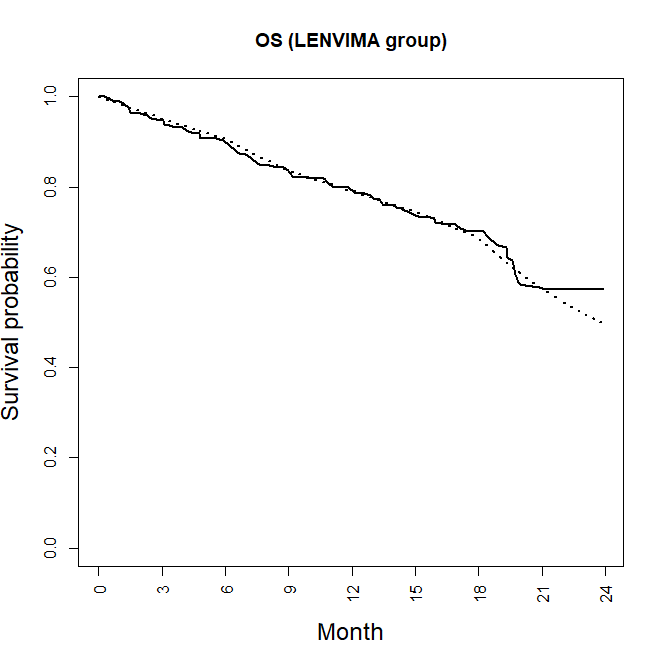}
 \end{subfigure}

 \caption{Fitted PFS and OS curves, where the solid lines are re-engineered survival curves from the originally reported K-M curves, and the dotted lines are based on three-state models fit to the re-engineered data.}
 \label{fig:Fitted}

\end{figure}

The review document for SELECT provides the Kaplan-Meier (K-M) curves for progression-free survival (PFS), overall survival (OS), and survival after crossover \citep{SELECT2019}. Based on these curves, we used the \textsf{R} package \texttt{IPDfromKM} (version 0.1.10 on CRAN) to reconstruct the individual patient data. We then used piecewise exponential models to estimate the hazard rates. We determined the three underlying hazard functions in the three-state models (for treatment and control separately) through trial and error, so that the resulting PFS and OS curves largely matched the reported K-M curves in SELECT. Figure \ref{fig:Fitted} plots the re-engineered survival curves from the originally reported K-M curves (solid lines) vs. the fitted PFS and OS curves (dotted lines). 

If we believe that the estimated parameters reflect the underlying truth, then we can design a simulation study based on these parameters. In our simulation study, we assumed that the recruitment was completed in 6 months with a uniform accrual rate. Out of the 392 patients, 261 were randomized to LENVIMA and 131 to placebo (roughly a 2:1 ratio). After randomization, patients were followed to the study cut-off at 24 months or to the censoring time, which was assumed to follow an exponential distribution with monthly hazard rate of $0.004$. {\revise We assumed that $83\%$ patients in the placebo group switched to open-label LENVIMA while the rest remained in the placebo group (i.e. $\pi_2 = 0.83$). It should be noted that the actual LENVIMA trial had 100\% crossover for the placebo patients. However, if we were to assume $\pi_2=1$, then we would not be able to match the PFS and OS curves without having access to individual data on when crossover occurred. Thus, we set $\pi_2=0.83$ in our study.}  

Under the settings of our simulation, the HR based on the ITT analysis was $0.859$ favoring the treatment group, and the power was $15.7\%$ based on a two-sided log-rank test with significance level $\alpha=0.05$. However, if placebo patients were \emph{not} allowed to switch to open-label LENVIMA after disease progression, then the hazard ratio would drop to $0.236$. This is the true treatment effect when crossover is \emph{not} allowed. The power in this case would be $100\%$. As before, these results were obtained using the \textsf{R} package \texttt{PWEALL} (version 1.4.0 on CRAN). 

\begin{figure}\label{s}
    \centering
    \includegraphics[width=12cm,height=8cm]{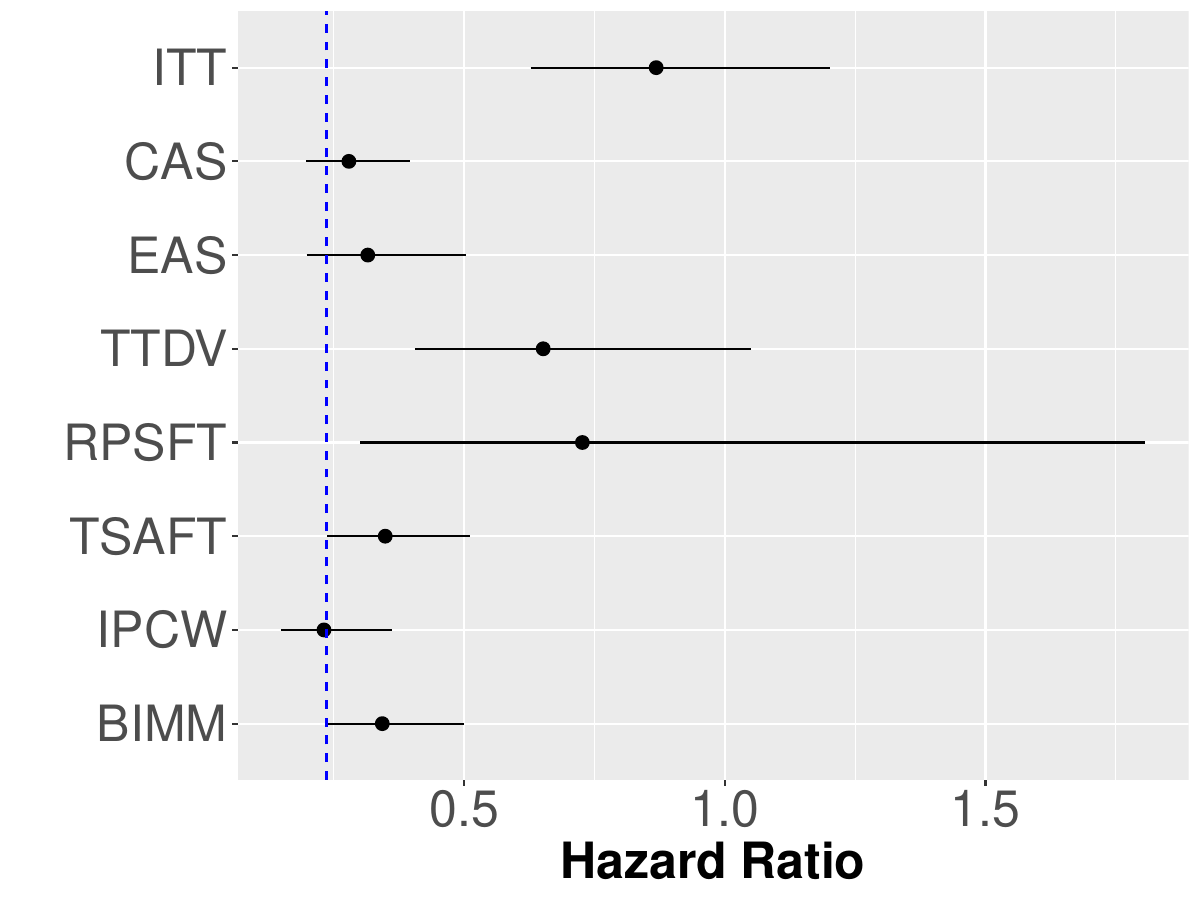}
    \caption{Simulation results from the reverse-engineered LENVIMA data. The dotted blue line denotes true treatment effect when crossover is not allowed.}
    \label{fig:rekm}
\end{figure}

Our simulation results averaged across 2000 replicates are plotted in Figure \ref{fig:rekm}. A lower estimated treatment effect indicates a larger discrepancy between the LENVIMA and the placebo groups, and thus, suggests greater efficacy of the drug. Figure \ref{fig:rekm} shows that the estimated HR from the unadjusted ITT analysis deviated significantly from the true HR. After adjusting for crossover, all other methods concluded greater drug effectiveness than that determined by the ITT analysis. 

Figure \ref{fig:rekm} also shows that the PP methods (CAS and EAS) and the IPCW method performed the best on average, with the lowest magnitude of average bias and the highest empirical coverage. In this particular experiment with reengineered data from the LENVIMA trial, the treatment effect could be best estimated using only
data before crossover.
{\revise This could possibly be due to the Markov crossover assumption \eqref{Markov_assumption} being reasonable in this experimental setting. The Markov crossover assumption is suitable, for example, in situations where the treatment has a relatively short half-life \citep{xiaodong2019}. In this scenario, the hazard rate may not change for patients who have switched from the control group to the experimental treatment group, and therefore, it is appropriate to only use the data from before crossover. However, we caution that this is just one synthetic experiment meant to mimic the LENVIMA trial. On real clinical data, researchers should assess the appropriateness of excluding or censoring data after crossover, as well as the implicit assumptions made by the CAS, EAS, or IPCW methods.}


\section{Discussion} \label{Sec:Discussion}

We have proposed the TSM framework which unifies existing and new methods for treatment effect estimation in RCTs with treatment crossover. A number of existing methods (Table \ref{tab:summary}) are covered in this framework, which allows delineation of these methods in terms of underlying assumptions and limitations. {\revise This allows practitioners to better understand each method's assumptions and choose the best approach for their own clinical data}. The TSM framework also allows for the creation of \emph{new} methods that incorporate diverse assumptions tailored to specific scenarios, thus enhancing its applicability in RCTs. A new imputation method for modeling the counterfactual survival time under treatment crossover was proposed and illustrated under this framework. By adopting a statistical perspective, our work complements and augments the comparative numerical studies of different methods by \cite{Morden2011} and \cite{Latimer2018}.

Modeling the data after treatment crossover to estimate the true treatment effect is essentially an imputation problem. In general, post-crossover data are biased, and different methods treat the biased data in different ways. The best choice of imputation method depends on {\revise the context of the RCT, the nature of the data, and the practitioners' assumptions about the type of crossover}.
This is especially the case when $100\%$ of patients in the control group switch to the experimental treatment, and thus, the true treatment effect is not identifiable without more stringent assumptions. 

Apart from the Markov \eqref{Markov_assumption} and semi-Markov \eqref{semi-Markov_assumption} crossover assumptions, other implicit assumptions for each method also need to be carefully assessed (e.g. the assumption of a constant treatment effect in TTDV and RPSFT, {\revise the no confounding assumption in TSAFT and IPCW}, the assumption of conditional independence of time-to-event and time-to-crossover given covariates in IPCW, etc.). It is important for researchers to be transparent about potential sources of bias inherent in their assumptions and properly report the limitations of their approaches. {\revise In clinical trials, extensive simulation studies and assumption checking (whenever possible) are both recommended to confirm the appropriateness of their chosen methods.}

\section*{Acknowledgments}

We are grateful to the three anonymous reviewers whose thoughtful feedback helped us greatly improve our paper. 
\bibliographystyle{agsm}

\bibliography{Bibliography-MM-MC}

\begin{appendix}

\section{{\revise Random number generation in the TSM}}
{\revise
Based on the TSM framework and the additional assumption $P(U>T)=0$, we can write the conditional hazard function of $T$ given $U=u$ as 
\begin{eqnarray*}
\lambda(t\mid u)=\lambda_{1}(t)I(t\le u)+\lambda^{{\bf x}}_{2}(t\mid u)I(t>u).
\end{eqnarray*}
The univariate function $\lambda_1(t)$ is the hazard rate without crossover and the bivariate function $\lambda^{{\bf x}}_{2}(t\mid u)$ is the hazard rate from crossover to event. This second hazard rate may depend on when the crossover occurs. Note that the joint distribution can be viewed as being conditional on $U \le T$, and $T$ and $U$ are not assumed to be independent of each other. 

\begin{description}
\item[Markov crossover.] When $\lambda^{\bf x}_2(t\mid u)=\lambda_2(t)$, the survival function $\widetilde{S}(t)$ of the event time $T$ is
\begin{eqnarray*}
\widetilde{S}(t)=P(T>t)=S_1(t)S_3(t)+\int_0^t \frac{S_1(u)S_3(u)S_2(t)}{S_2(u)}\lambda_3(u)du,
\end{eqnarray*}
where $S_k(t)=\exp\{-\int_0^t \lambda_k(u)du\}$, $k=1,2,3$. To generate $(T,U)$, we use the following steps:
\begin{description}
\item[(MC1)] Generate two independent random variables $X$ and $U$ such that $X$ follows $\mbox{Unif}(0,1)$ and $U$ follows $F_3=1-S_3$.
\item[(MC2)] If $X>S_1(U)$, set $T=F_1^{-1}(1-X)$; else set $T=F_2^{-1}\{1-XS_2(U)/S_1(U)\}$, where $F_k^{-1}$ is the inverse function of $F_k=1-S_k$, $k=1,2$. 
\end{description}

\item[Semi-Markov crossover.] When $\lambda^{\bf x}_2(t\mid u)=\lambda_2(t-u)$, the marginal survival function $\widetilde{S}(t)$ of the event time is
\begin{eqnarray*}
\widetilde{S}(t)=P(T>t)=S_1(t)S_3(t)+\int_0^t S_1(u)S_3(u)S_2(t-u)\lambda_3(u)du,
\end{eqnarray*}
where $S_k(t)=\exp\{-\int_0^t \lambda_k(u)du\}$, $k=1,2,3$. To generate $(T,U)$, we use the following steps:
\begin{description}
\item[(SMC1)] Generate two independent random variables $X$ and $U$ such that $X$ follows $\mbox{Unif}(0,1)$ and $U$ follows $F_3=1-S_3$.
\item[(SMC2)] If $X>S_1(U)$, set $T=F_1^{-1}(1-X)$; else set $T=U+F_2^{-1}\{1-X/S_1(U)\}$, where $F_k^{-1}$ is the inverse function of $F_k=1-S_k$, $k=1,2$.
\end{description}

\item[General crossover.] For general $\lambda^{\bf x}_2(t\mid u)$, the marginal survival function,
\begin{eqnarray*}
\widetilde{S}(t)=P(T>t)=S_1(t)S_3(t)+\int_0^t {S_1(u)S_3(u)S_{2}(t\mid u)}\lambda_3(u)du,
\end{eqnarray*}
where $S_{2}(t\mid u)=\exp\{-\int_u^t \lambda^{\bf x}_2(s\mid u)ds\}$. We use the following steps:
\begin{description}
\item[(GC1)] Generate two independent random variables $X$ and $U$ such that $X$ follows $\mbox{Unif}(0,1)$ and $U$ follows $F_3=1-S_3$.
\item[(GC2)] If $X>S_1(U)$, set $T=F_1^{-1}(1-X)$, where $F_1^{-1}$ is the inverse function of $F_1=1-S_1$.
\item[(GC3)] If $X\le S_1(U)$, set 
$$T=F_{2,U}^{-1}\{1-X/S_1(U)\},$$
where, $S_{2,u}(t)=S_2(t\mid u)$, $F_{2,u}=1-S_{2,u}$ and $F_{2,u}^{-1}$ is the inverse function of $F_{2,u}$ for any fixed $u$. 
\end{description}

\end{description}

\noindent This approach to generating the crossover time and event time in TSM (assuming that all the pertaining hazard functions are piecewise constant functions) has been implemented using the function \texttt{rpwecx} in the \textsf{R} package \texttt{PWEALL} (version 1.4.0 on CRAN). }

\section{{\revise Additional Tables and Figures}} \label{App:A}

\begin{figure}[t]
\begin{subfigure}[h]{0.5\linewidth}
\includegraphics[width=\linewidth]{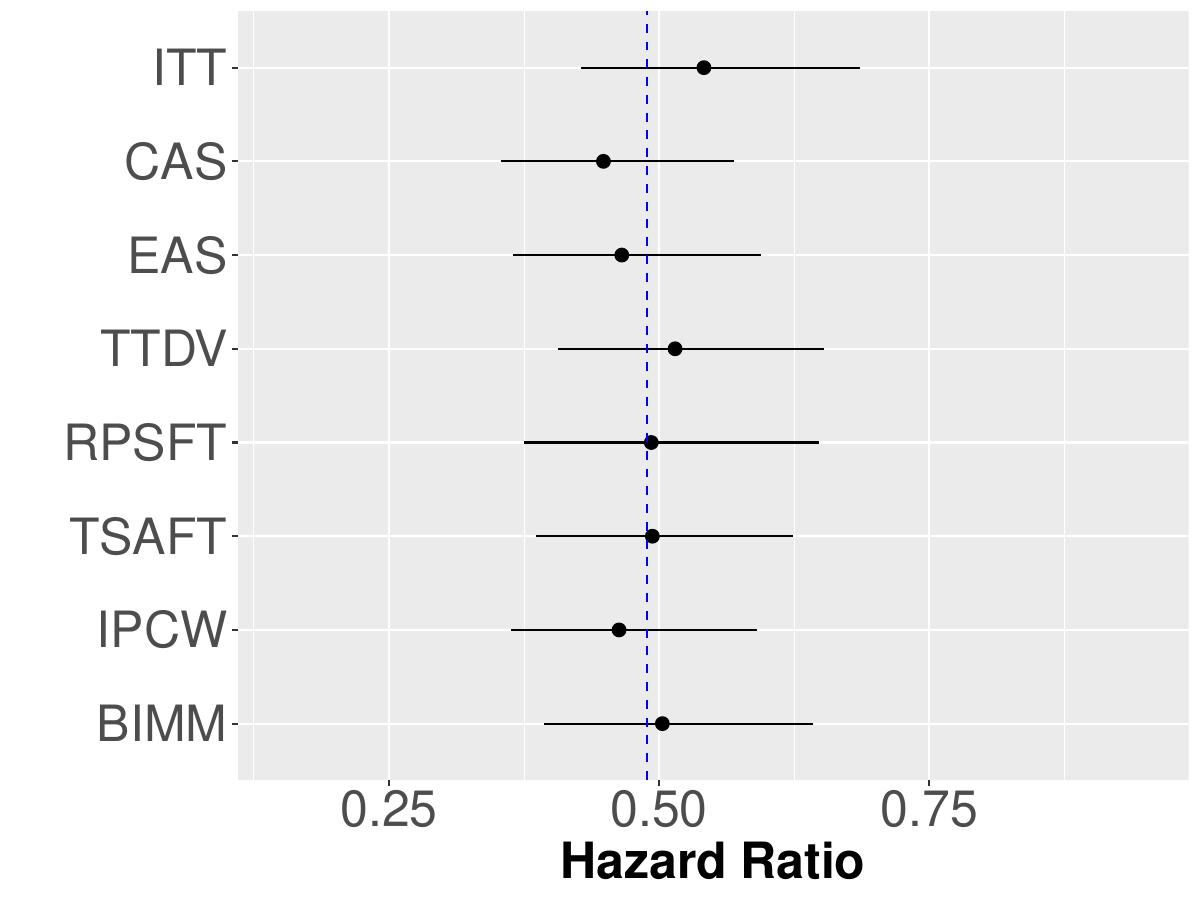}
\caption{25\% crossover after progression}
\end{subfigure}
\hfill
\begin{subfigure}[h]{0.5\linewidth}
\includegraphics[width=\linewidth]{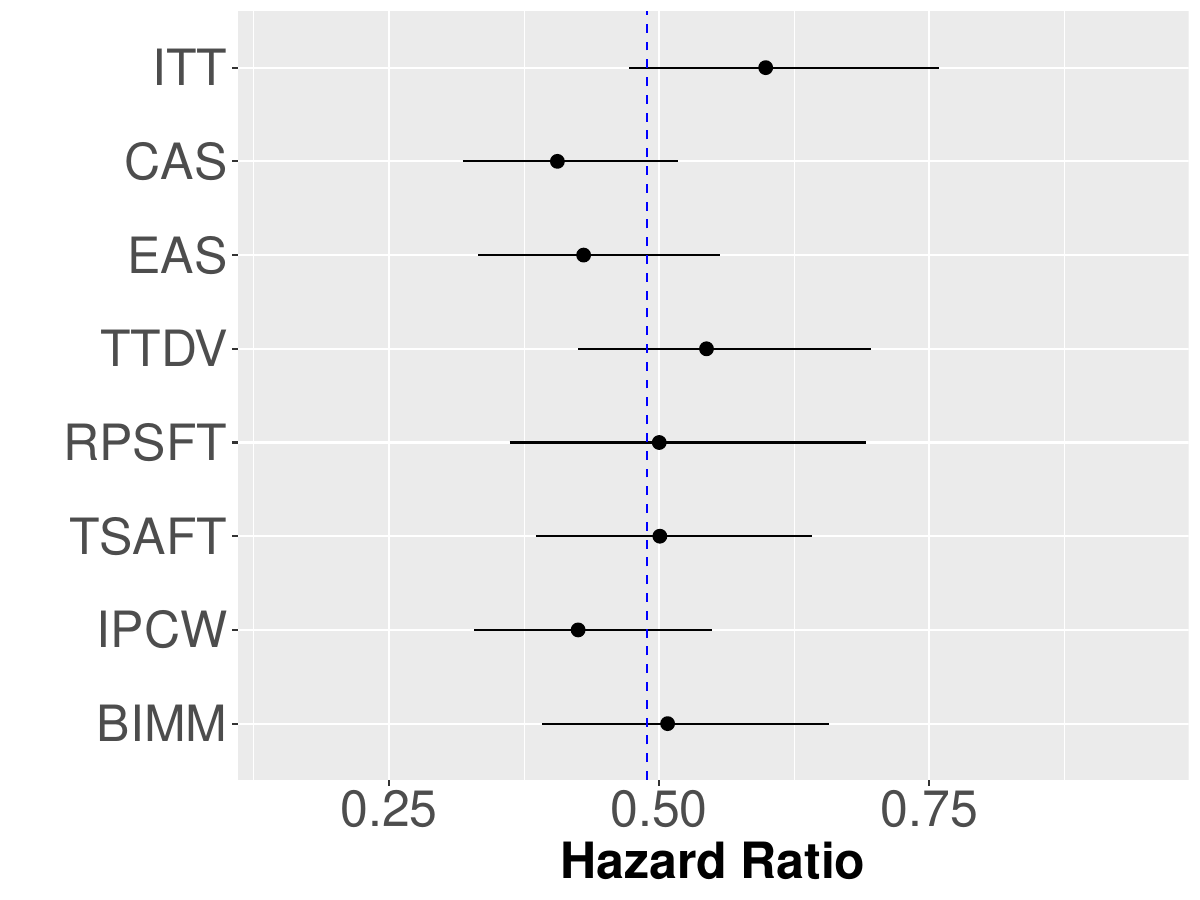}
\caption{50\% crossover after progression}
\end{subfigure}%
 \\[.25in]
\begin{subfigure}[h]{0.5\linewidth}
\includegraphics[width=\linewidth]{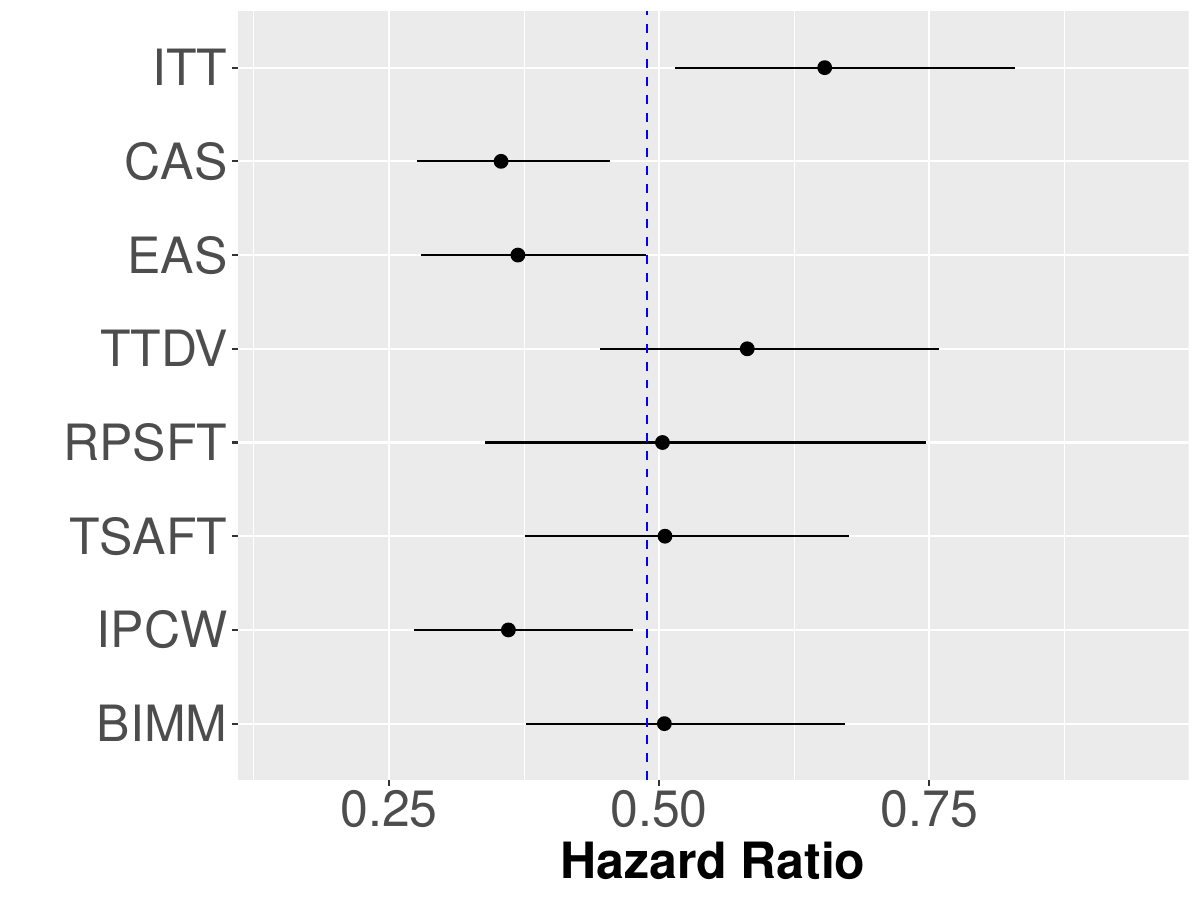}
\caption{75\% crossover after progression}
\end{subfigure}
\hfill
\begin{subfigure}[h]{0.5\linewidth}
\includegraphics[width=\linewidth]{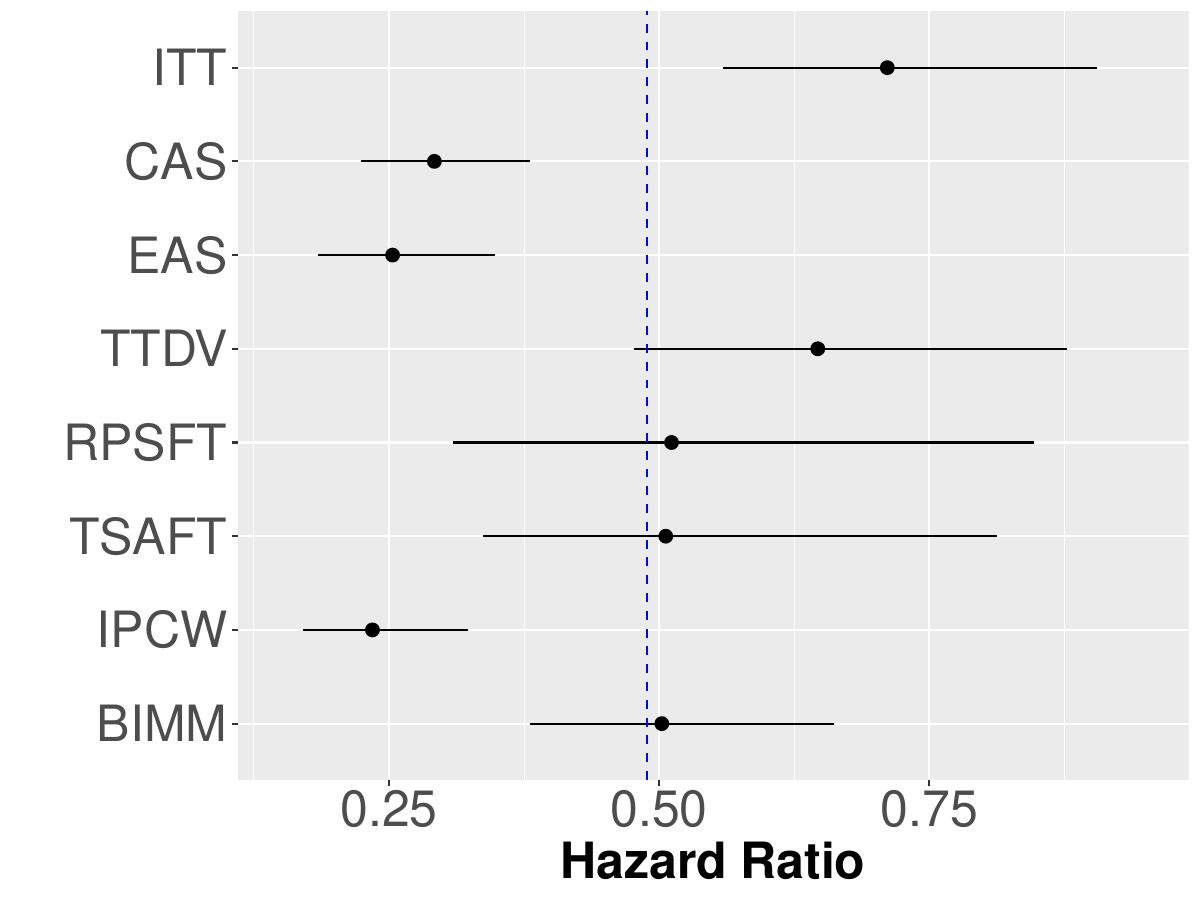}
\caption{100\% crossover after progression}
\end{subfigure}%
\caption{{\revise Simulation results under lower censoring  averaged across 2000 replicates. The dotted blue line is the true treatment effect ($\text{HR}_{\text{true}}=0.489$).}} \label{fig:low_censoring_simulation_results}
\end{figure}

\begin{figure}[t]
\begin{subfigure}[h]{0.5\linewidth}
\includegraphics[width=\linewidth]{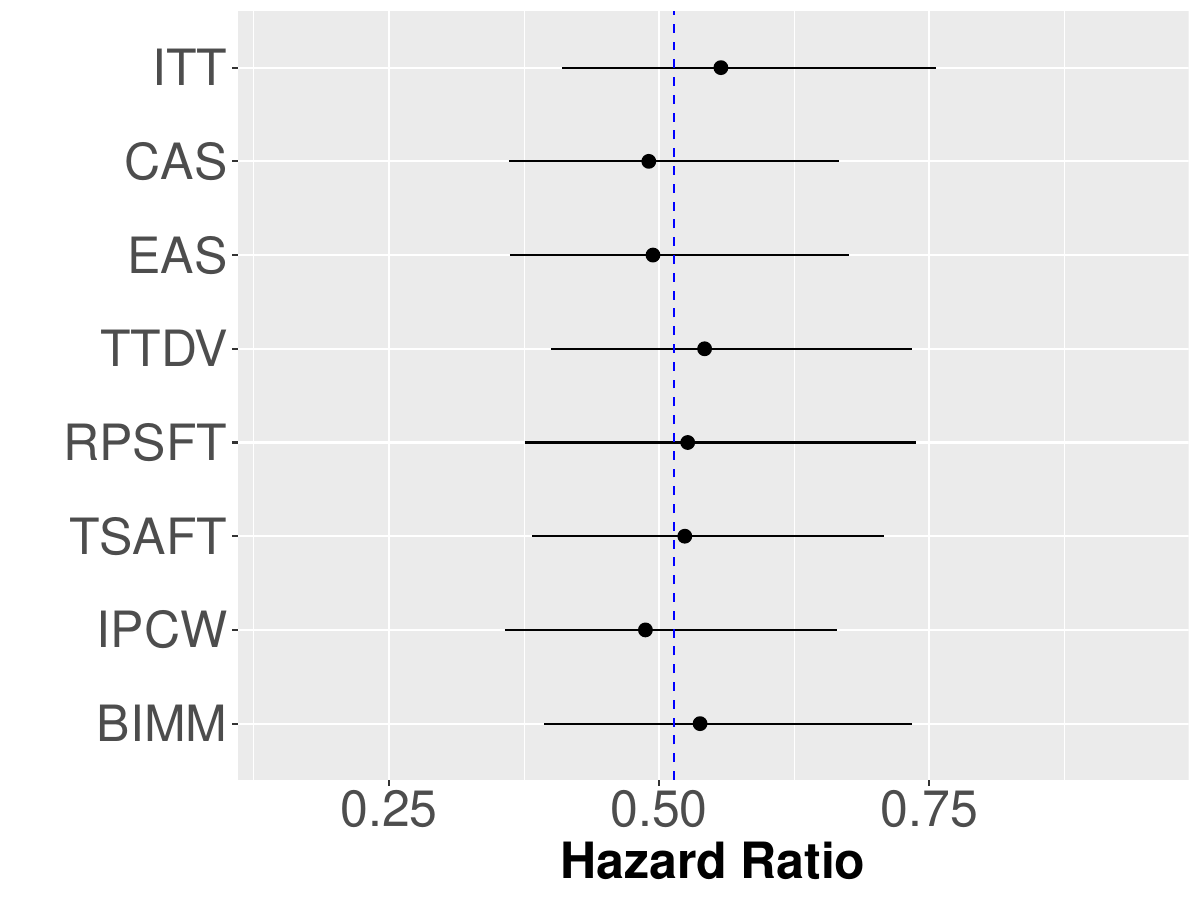}
\caption{25\% crossover after progression}
\end{subfigure}
\hfill
\begin{subfigure}[h]{0.5\linewidth}
\includegraphics[width=\linewidth]{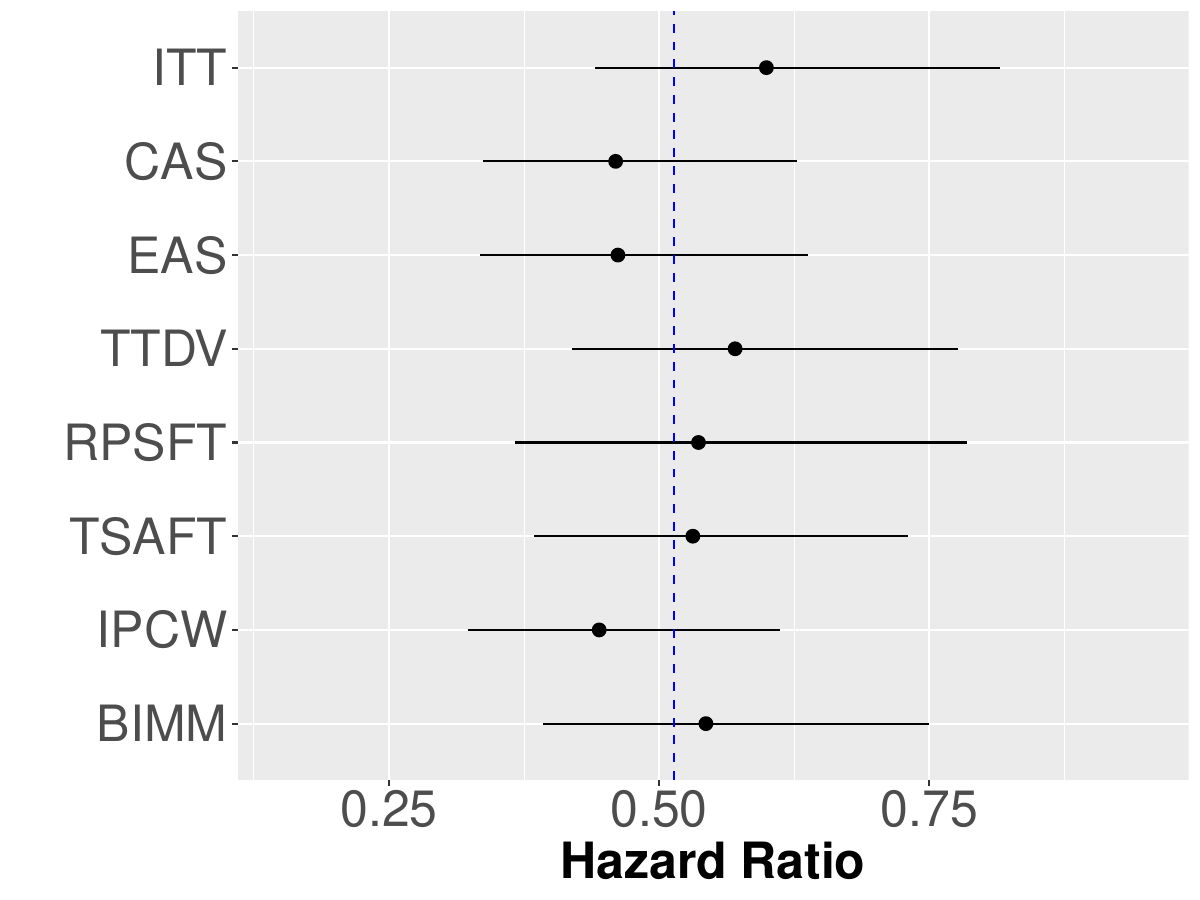}
\caption{50\% crossover after progression}
\end{subfigure}
\\[.25in]
\begin{subfigure}[h]{0.5\linewidth}
\includegraphics[width=\linewidth]{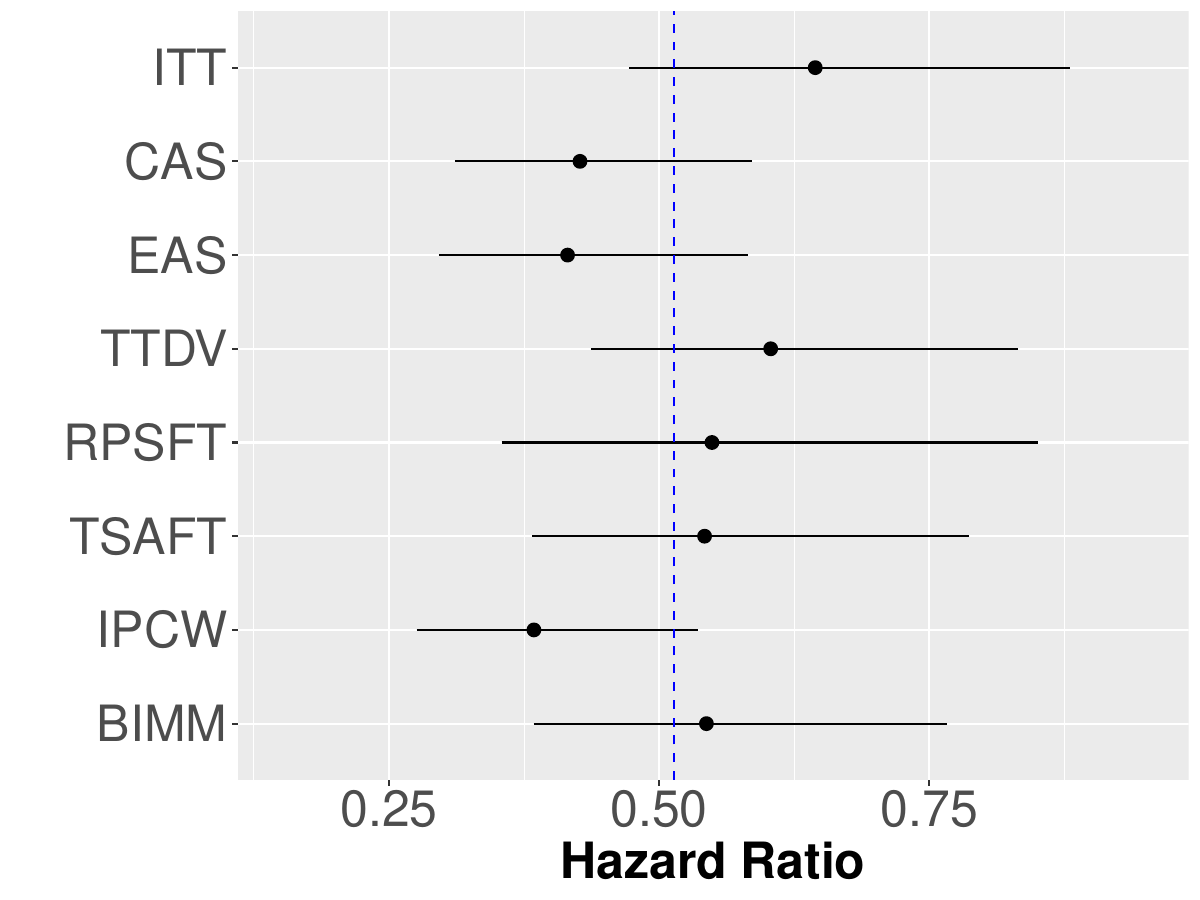}
\caption{75\% crossover after progression}
\end{subfigure}
\hfill
\begin{subfigure}[h]{0.5\linewidth}
\includegraphics[width=\linewidth]{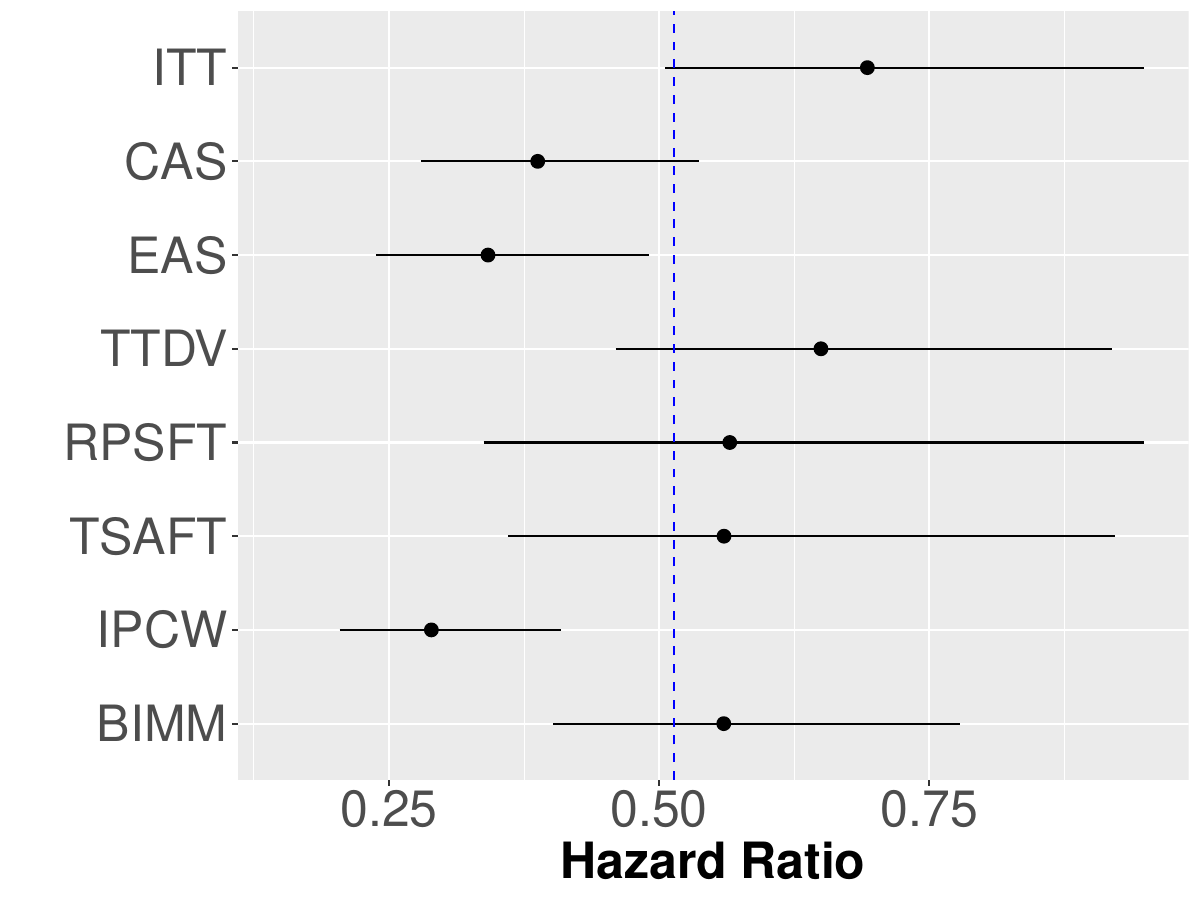}
\caption{100\% crossover after progression}
\end{subfigure}%
\caption{{\revise Simulation results under higher censoring averaged across 2000 replicates. The dotted blue line is the true treatment effect ($\text{HR}_{\text{true}}=0.514$).}} \label{fig:high_censoring_simulation_results}
\end{figure}

{\revise In this section, we examine the effect of the censoring rate on the different TSM methods. We repeated the experiments from Section \ref{subsec:ex1} under lower and higher censoring rates than those of the simulations in Section \ref{subsec:ex1}. Our experiments in Section \ref{subsec:ex1} had approximately 38.4\%, 40\%, 41.6\% and 43.2\% censored patients respectively when $\pi_2$ = 0.25, 0.5, 0.75 and 1. We deem these scenarios as ``moderate censoring.''

To attain lower censoring rates, we assumed that the yearly censoring rate was $0.02$. The study had a one-year recruitment period with a uniform accrual rate to recruit $400$ patients, and the study would be read out at 8 years after the randomization of the first patient. This resulted in roughly 28.4\%, 29.9\%, 31.4\% and 32.9\% censored patients respectively when $\pi_2$ = 0.25, 0.5, 0.75 and 1.

To attain higher censoring rates, we assumed that the yearly censoring rate was $0.025$. The study had a two-year recruitment period with a uniform accrual rate to recruit $400$ patients, and the study would be read out at 5 years after the randomization of the first patient. This resulted in roughly 50.5\%, 51.9\%, 53.2\% and 54.6\% censored patients respectively when $\pi_2$ = 0.25, 0.5, 0.75 and 1.

The results for lower censoring are shown in Figure \ref{fig:low_censoring_simulation_results} and Table \ref{tab:low_censoring_crossover_results}, while the results for higher censoring are shown in Figure \ref{fig:high_censoring_simulation_results} and Table \ref{tab:high_censoring_crossover_results}. Overall, the average bias and SE for all methods tended to be slightly lower under lower censoring than moderate censoring. Meanwhile, average bias and MSE tended to be higher under higher censoring. This is what we expect since higher censoring percentages inherently make estimation of the true treatment effect more difficult. Interestingly, however, the censoring percentage did not have as much of an impact on the ECP. This could be because higher amounts of censoring also tended to lead to higher SE's (or greater uncertainty) and therefore more conservative 95\% CIs.

Regardless of lower, moderate, or higher censoring rates, our results in Figures \ref{fig:low_censoring_simulation_results} and \ref{fig:high_censoring_simulation_results} and Tables \ref{tab:low_censoring_crossover_results} and \ref{tab:high_censoring_crossover_results} are consistent with those in Section \ref{Sec:sims}. Namely,  the TSM methods that assumed semi-Markov crossover and that used the data after crossover (i.e. RPSFT, TSAFT, and BIMM) consistently outperformed the other methods (ITT, CAS, EAS, TTDV, and IPWC). Moreover, the performance of ITT, CAS, EAS, TTDV, and IPWC degraded considerably as the crossover propotion $\pi_2$ increased. This suggests that under many different scenarios with different censoring and crossover proportions, it may be prudent for practitioners to consider using the data after crossover to estimate treatment effects.}

\begin{table}[ht]
\centering
\captionof{table}{{\revise Simulation results under lower censoring based on 2000 replications. The table displays the average Bias, SE, and MSE from all Monte Carlo replicates, while the ECP is the percentage of 95\% CIs that contained the true HR.}}
\begin{tabular}{l c cccc c cccc}  
\toprule
 & & \multicolumn{4}{c}{\textbf{25\% crossover}} & & \multicolumn{4}{c}{\textbf{50\% crossover}} \\
\cmidrule(lr){3-6} \cmidrule(lr){8-11}
\textbf{Method} & & Bias & SE & MSE & ECP (\%) & & Bias & SE & MSE & ECP (\%) \\
\midrule
ITT & & 0.052 & 0.065 & 0.011 & 87.1 & & 0.109 & 0.072 & 0.023 & 63.9 \\
CAS & & -0.040 & 0.054 & 0.007 & 87.6 & & -0.083 & 0.050 & 0.012 & 65.4 \\
EAS & & -0.024 & 0.058 & 0.007 & 92.2 & & -0.058 & 0.056 & 0.010 & 82.7\\
TTDV & & 0.025 & 0.062 & 0.008 & 92.9 & & 0.054 & 0.063 & 0.012 & 87.3 \\
RPSFT & & 0.003 & 0.075 & 0.010 & 96.0 & & 0.012 & 0.090 & 0.014 & 96.0 \\
TSAFT & & 0.004 & 0.062 & 0.008 & 93.3 & & 0.011 & 0.066 & 0.009 & 93.3 \\
IPCW & & -0.025 & 0.057 & 0.007 & 92.2 & & -0.063 & 0.055 & 0.010 & 81.0 \\
BIMM & & 0.014 & 0.062 & 0.008 & 94.0 &  & 0.018 & 0.066 & 0.009 & 94.3 \\
\end{tabular} 
\\[.1in]
\begin{tabular}{l c cccc c cccc}  
 & & \multicolumn{4}{c}{\textbf{75\% crossover}} & & \multicolumn{4}{c}{\textbf{100\% crossover}} \\
\cmidrule(lr){3-6} \cmidrule(lr){8-11}
\textbf{Method} & & Bias & SE & MSE & ECP (\%) & & Bias & SE & MSE & ECP (\%) \\
\midrule
ITT & & 0.164 & 0.079 & 0.039 & 36.9 & &0.222 & 0.087 & 0.065 & 16.2 \\
CAS & & -0.135 & 0.045 & 0.022 & 26.2 & & -0.196 & 0.039 & 0.042 & 1.9 \\
EAS & & -0.119 & 0.052 & 0.020 & 47.4 & & -0.235 & 0.041 & 0.059 & 1.6 \\
TTDV & & 0.092 & 0.078 & 0.021 & 76.6 & & 0.157 & 0.100 & 0.046 & 60.1 \\
RPSFT & & 0.014 & 0.109 & 0.021 & 97.5 & & 0.022 & 0.140 & 0.034 & 98.9 \\
TSAFT & & 0.016 & 0.078 & 0.012 & 94.0 & & 0.017 & 0.125 & 0.031 & 93.5 \\
IPCW & & -0.128 & 0.051 & 0.021 & 39.3 & & -0.254 & 0.038 & 0.067 & 0.15 \\
BIMM & & 0.015 & 0.074 & 0.011 & 95.0 &  & 0.013 & 0.070 & 0.019 & 77.1 \\
\bottomrule
\end{tabular} \label{tab:low_censoring_crossover_results}
\end{table}

\newpage 

\begin{table}[t]
\centering
\captionof{table}{{\revise Simulation results under higher censoring based on 2000 replications. The table displays the average Bias, SE, and MSE from all Monte Carlo replicates, while the ECP is the percentage of 95\% CIs that contained the true HR.}}
\begin{tabular}{l c cccc c cccc}  
\toprule
 & & \multicolumn{4}{c}{\textbf{25\% crossover}} & & \multicolumn{4}{c}{\textbf{50\% crossover}} \\
\cmidrule(lr){3-6} \cmidrule(lr){8-11}
\textbf{Method} & & Bias & SE & MSE & ECP (\%) & & Bias & SE & MSE & ECP (\%) \\
\midrule
ITT & & 0.042 & 0.086 & 0.017 & 91.8 & & 0.084 & 0.094 & 0.025 & 83.9 \\
CAS & & -0.024 & 0.076 & 0.013 & 92.8 & & -0.054 & 0.072 & 0.014 & 87.7 \\
EAS & & -0.020 & 0.078 & 0.013 & 93.3 & & -0.052 & 0.076 & 0.015 & 87.6 \\
TTDV & & 0.027 & 0.083 & 0.015 & 93.9 & & 0.055 & 0.089 & 0.020 & 89.8 \\
RPSFT & & 0.011 & 0.099 & 0.018 & 95.4 & & 0.021 & 0.112 & 0.023 & 96.2 \\
TSAFT & & 0.009 & 0.085 & 0.015 & 93.9 & & 0.016 & 0.090 & 0.017 & 93.7 \\
IPCW & & -0.027 & 0.077 & 0.013 & 92.7 & & -0.070 & 0.072 & 0.015 & 83.7 \\
BIMM & & 0.023 & 0.085 & 0.015 & 93.7 &  & 0.028 & 0.089 & 0.017 & 94.0 \\
\end{tabular} 
\\[.1in]
\begin{tabular}{l c cccc c cccc}  
 & & \multicolumn{4}{c}{\textbf{75\% crossover}} & & \multicolumn{4}{c}{\textbf{100\% crossover}} \\
\cmidrule(lr){3-6} \cmidrule(lr){8-11}
\textbf{Method} & & Bias & SE & MSE & ECP (\%) & & Bias & SE & MSE & ECP (\%) \\
\midrule
ITT & & 0.129 & 0.102 & 0.038 & 73.2 & &0.178 & 0.111 & 0.056 & 56.4 \\
CAS & & -0.087 & 0.068 & 0.017 & 76.6 & & -0.126 & 0.064 & 0.024 & 59.0 \\
EAS & & -0.099 & 0.071 & 0.020 & 73.9 & & -0.172 & 0.063 & 0.038 & 37.4 \\
TTDV & & 0.088 & 0.099 & 0.027 & 85.6 & & 0.135 & 0.114 & 0.045 & 76.4 \\
RPSFT & & 0.034 & 0.131 & 0.032 & 97.1 & & 0.050 & 0.157 & 0.047 & 98.1 \\
TSAFT & & 0.027 & 0.106 & 0.022 & 94.3 & & -0.045 & 0.148 & 0.046 & 94.0 \\
IPCW & & -0.130 & 0.065 & 0.025 & 55.4 & & -0.225 & 0.051 & 0.055 & 6.25 \\
BIMM & & 0.029 & 0.095 & 0.019 & 95.0 &  & 0.045 & 0.094 & 0.030 & 82.8 \\
\bottomrule
\end{tabular} \label{tab:high_censoring_crossover_results}
\end{table}

\end{appendix}
\end{document}